\PassOptionsToPackage{table}{xcolor}
\documentclass[sigconf]{acmart}
\AtBeginDocument{%
  }

\usepackage{xspace}
\usepackage{makecell}
\usepackage{tabularx}
\usepackage{enumitem}
\usepackage{float}
\usepackage{fvextra}
\usepackage{fancyvrb}
\usepackage{soul}

\newcolumntype{L}{>{\raggedright\arraybackslash}X} 
\newcommand{\tool}{\textsc{TheraTrack}\xspace}
\renewcommand{\hl}[1]{#1}

\copyrightyear{2026}
\acmYear{2026}
\setcopyright{cc}
\setcctype{by}
\acmConference[CHI '26]{Proceedings of the 2026 CHI Conference on Human Factors in Computing Systems}{April 13--17, 2026}{Barcelona, Spain}
\acmBooktitle{Proceedings of the 2026 CHI Conference on Human Factors in Computing Systems (CHI '26), April 13--17, 2026, Barcelona, Spain}
\acmPrice{}
\acmDOI{10.1145/3772318.3790569}
\acmISBN{979-8-4007-2278-3/2026/04}

\begin{document}

\title{Exploring Customizable Interactive Tools for Therapeutic Homework Support in Mental Health Counseling}


\author{Yimeng Wang}  
\affiliation{
    \department{Computer Science}
    \institution{William \& Mary}
    \city{Williamsburg}
    \state{VA}
    \country{USA}}
\email{ywang139@wm.edu}
\orcid{0009-0005-0699-4581} 

\author{Liabette Escamilla}  
\affiliation{
    \department{Counseling Center}
    \institution{William \& Mary}
    \city{Williamsburg}
    \state{VA}
    \country{USA}}
\email{laescamilla@wm.edu}
\orcid{0000-0003-1233-0376}

\author{Yinzhou Wang}  
\affiliation{
    \department{Computer Science}
    \institution{William \& Mary}
    \city{Williamsburg}
    \state{VA}
    \country{USA}}
\email{ywang143@wm.edu}
\orcid{0009-0009-6355-2551} 

\author{Bianca R. Augustine}  
\affiliation{
    \department{School Psychology \& Counselor Education}
    \institution{William \& Mary}
    \city{Williamsburg}
    \state{VA}
    \country{USA}}
\email{braugustine@wm.edu}
\orcid{0000-0002-4313-8283} 

\author{Yixuan Zhang}  
\affiliation{
    \department{Computer Science}
    \institution{William \& Mary}
    \city{Williamsburg}
    \state{VA}
    \country{USA}} 
\email{yzhang104@wm.edu}
\orcid{0000-0002-7412-4669}

\renewcommand{\shortauthors}{Wang et al.}

\begin{abstract}
Therapeutic homework (i.e., tasks assigned by therapists for clients to complete between sessions) is essential for effective psychotherapy, yet therapists often interpret fragmented client logs, assessments, and reflections within limited preparation time. Our formative study with licensed therapists revealed three critical design requirements: support for interpreting unstructured client self-reports, customization aligned with clinical objectives, and seamless integration across multiple data sources. We then designed and developed \tool, a customizable, therapist-facing tool that integrates multi-dimensional data and leverages large language models to generate traceable summaries and support natural-language queries, to streamline between-session homework tracking. Our pilot study with 14 therapists showed that \tool reduced their cognitive load, enabled verification through direct navigation from AI summaries to original data entries, and was adapted differently for private analysis compared to in-session use, with dependence varying based on therapist experience and usage duration. We also discuss design implications for clinician-centered AI for mental health. 
\end{abstract}





\maketitle

\section{Introduction}

\begin{figure*}
\centering
\includegraphics[width=\linewidth]{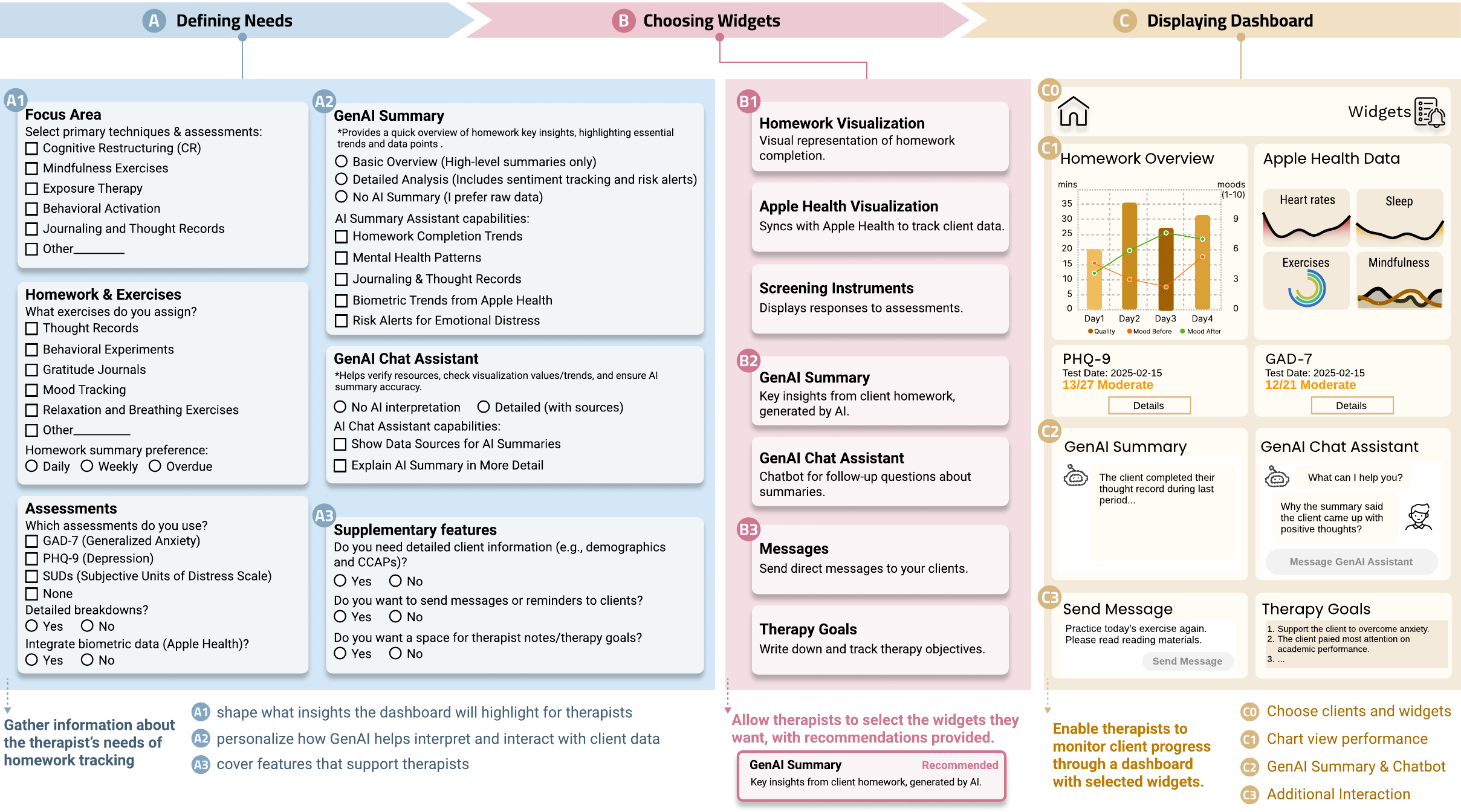}
\caption{Overview of the three-step flow for configuring a therapist-facing tool \tool. Therapist-defined need in \includegraphics[height=3mm]{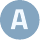} corresponds to specific widget options in \includegraphics[height=3mm]{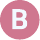}, which are then realized as interactive dashboard components in \includegraphics[height=3mm]{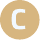}. \textbf{Defining Needs:} Therapists complete an onboarding survey to customize their preferences for tracking homework and client progress, such as \includegraphics[height=3mm]{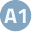} therapists' focus areas, homework types, and assessments, \includegraphics[height=3mm]{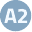} setting preferences for GenAI summaries and assistance, and \includegraphics[height=3mm]{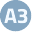} supplementary features including clinical information display and side functions. \textbf{Choosing Widgets:} Based on responses from step \includegraphics[height=3mm]{figures/A.pdf}, therapists are guided to select relevant \tool's widgets, such as \includegraphics[height=3mm]{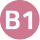} chart of homework progress and health, \includegraphics[height=3mm]{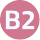} GenAI-generated summaries and chat assistant, and \includegraphics[height=3mm]{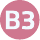} message and therapy goals. \textbf{Displaying Customized Dashboard:} A customized dashboard will then be generated with the selected widgets, enabling therapists to \includegraphics[height=3mm]{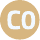} choose clients, adjust widgets, and access original homework submissions, \includegraphics[height=3mm]{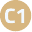} view homework trends and assessment results, \includegraphics[height=3mm]{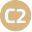} interact with GenAI summaries and chat assistants, and \includegraphics[height=3mm]{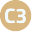} message to their clients and modify therapy goals.}
\label{fig:overview_of_dashboard}
\end{figure*}

In many forms of evidence-based psychotherapy, client ``homework'' plays a central role in supporting behavior change and reinforcing therapeutic insights between sessions~\cite{Mausbach2010, Kazantzis2010, Scheel2004, Horvath2011}. These assignments, such as journaling, mood tracking, or exercise worksheets, fall under the broader category of patient-generated health data (PGHD) that provides valuable insights into clients' daily experiences, patterns of thought, and struggles with change~\cite{Prasko2022}. In practice, therapists often receive homework assignments alongside supplementary clinical data (e.g., app-based emotional check-ins, standardized assessments), leaving therapists with the burden of piecing together fragmentary and heterogeneous information to maintain even a basic overview of client progress from one session to the next~\cite{Bunnell2024, Dattilio2011}. To facilitate the collection and organization of homework, a range of tools have emerged. Practice-management and measurement-based care infrastructures help collect standardized measures and distribute digital worksheets; client-facing mHealth tools scaffold journaling, mood tracking, and skills practice; and research prototypes explore digital workflows for assigning and retrieving homework~\cite{Bunnell2024, Dattilio2011}. These systems improve capture and access, but they rarely help therapists \emph{synthesize} heterogeneous inputs over time or audit how conclusions are reached.

For therapists, the challenge is not simply to read through these materials, but to critically review them to evaluate homework completion, interpret emerging patterns, situate shifts in content within the therapeutic trajectory, and determine how to respond in clinically appropriate ways~\cite{cully_dawson_hamer_tharp_2020}. To assist with review processes, some prototypes have explored dashboard visualizations of mood ratings to make fluctuations more visible, reports that automatically score assessments and generate brief rule-based summaries of results, and collaborative interfaces that allow therapists and clients to view journaling entries together, annotate them, or highlight notable events~\cite{grieg2019visual, Gondek2016, Reiter2022}. Meanwhile, recent advances in generative AI (GenAI)---particularly large language models (LLMs)---have raised the possibility of automating parts of review processes~\cite{Perttu2023}. LLMs can generate summaries, extract sentiment, and identify recurring themes from unstructured text, offering potential assistance in organizing fragmented client input. Recent work has explored the use of GenAI for therapy-adjacent tasks, such as progress note generation~\cite{Kambhamettu2024, Biswas2024, upheal, blueprint}, reflective journaling~\cite{Kim2024, Nepal2024}, and AI-mediated therapeutic conversations~\cite{Siddals2024, Heinz2025, Held2024}, demonstrating its promise to reduce clinician workload and broaden access to support. Yet existing work has largely examined whether such outputs are workable~\cite{Stade2024, Wang2025, FrankeFyen2025}---whether a generated note captures session details, whether a journaling reflection reads as coherent, or whether a chatbot provides sensible responses. What remains less understood is how therapists actually engage with these outputs, specifically, how they judge their relevance, adapt them to clinical goals, or calibrate trust in use.

Rather than treating design only as a path to technical solutions, we adopt a design-oriented research perspective~\cite{Fallman2003}, in which the process of prototyping becomes a means to interrogate how users' interpretive practices shape the very definition of ``support.'' This stance aligns with work in human–AI interaction and explainable AI, which provides design guidance for high-stakes settings, emphasizing the need to support appropriate reliance, align explanations with user questions, and make underlying sources auditable~\cite{amershi2019guidelines, liao2020questioning, buccinca2021trust}. Studies of human–AI collaboration further suggest that support is most effective when it adapts to users’ goals, within contexts, and calibrates trust appropriately~\cite{bansal2021, cai2019effects}. Yet little is known about how these principles translate to the context of psychotherapy homework support. 

In this work, we first conducted a formative survey study with 27 licensed therapists to identify challenges encountered in managing clients' between-session ``therapeutic homework''. \hl{Here, \textit{therapeutic homework} refers to tasks that therapists assign clients to complete outside of therapy sessions, including activities such as thought records, journaling, and behavioral experiments. While we focus primarily on these homework tasks, we also consider broader forms of PGHD, health-related information clients produce outside of formal clinical encounters, as therapists frequently interpret formal homework assignments alongside emotional check-ins, standardized assessments, and contextual details around significant events.} Therapists described cognitive burdens arising from manually synthesizing fragmented information, including emotional check-ins, thought monitoring, client self-reflections, and contextual details around significant events. Our results suggest three design requirements: 1) tools to support interpretation of clients' unstructured self-reflections; 2) customization capabilities aligning data presentation with specific clinical goals; and 3) a centralized platform integrating diverse data sources to streamline monitoring.

Based on the design requirements that emerged from our formative study, we designed and developed \tool, a therapist-facing application built around a customizable three-stage flow---\textit{Defining Needs}, \textit{Choosing Widgets}, and \textit{Displaying the Dashboard} (see \autoref{fig:overview_of_dashboard}). Therapists first complete a survey detailing their therapeutic focus areas, preferred clinical assessments, and desired depth of AI-generated analyses. Based on these preferences, \tool recommends relevant modular widgets, such as ``Homework Overview'', ``Assessment Trackers'', and ``GenAI-powered summaries''. Therapists can then customize their dashboard by selecting and arranging these widgets to best suit their workflow. The resulting dashboard integrates trends from formal assessments, detailed homework engagement analytics, and biometric data (e.g., sleep patterns, heart rate) from connected devices. Furthermore, a GenAI Summary widget synthesizes client journal entries into summaries, while a GenAI Chat Assistant enables therapists to pose specific, natural-language queries about client data.

We then evaluated \tool with 14 therapists. Participants reported cognitive relief through the system's ability to automate data synthesis, allowing them to focus on higher-level clinical interpretation. Therapists also valued how the integrated view helped them uncover nuanced insights they might have otherwise missed. The linkage of AI-generated insights to original data sources enhanced their trust and facilitated active verification. We also observed distinct patterns of use, with therapists customizing their dashboards differently for private analysis versus in-session client interactions, indicating \tool's flexibility and adaptability to clinical workflows. 

\textbf{Contributions:} We contribute:  
1) the characterization of therapists' needs and challenges regarding the management and sense-making of client therapeutic homework data; 
2) the design and development of an AI-supported therapeutic homework tracking tool \tool, a configurable and customizable application that integrates multi-dimensional data to support therapist sense-making of between-session homework; and 
3) a pilot study with therapists to assess the usability, perceived usefulness, and trust perceptions of \tool, shedding light on future design implications of AI-supported therapist-facing tools.  
   
\section{Background \& Related Work}
\subsection{General Background of Therapeutic Homework}
\begin{figure*}[h!]
    \centering
    \includegraphics[width=0.8\linewidth]{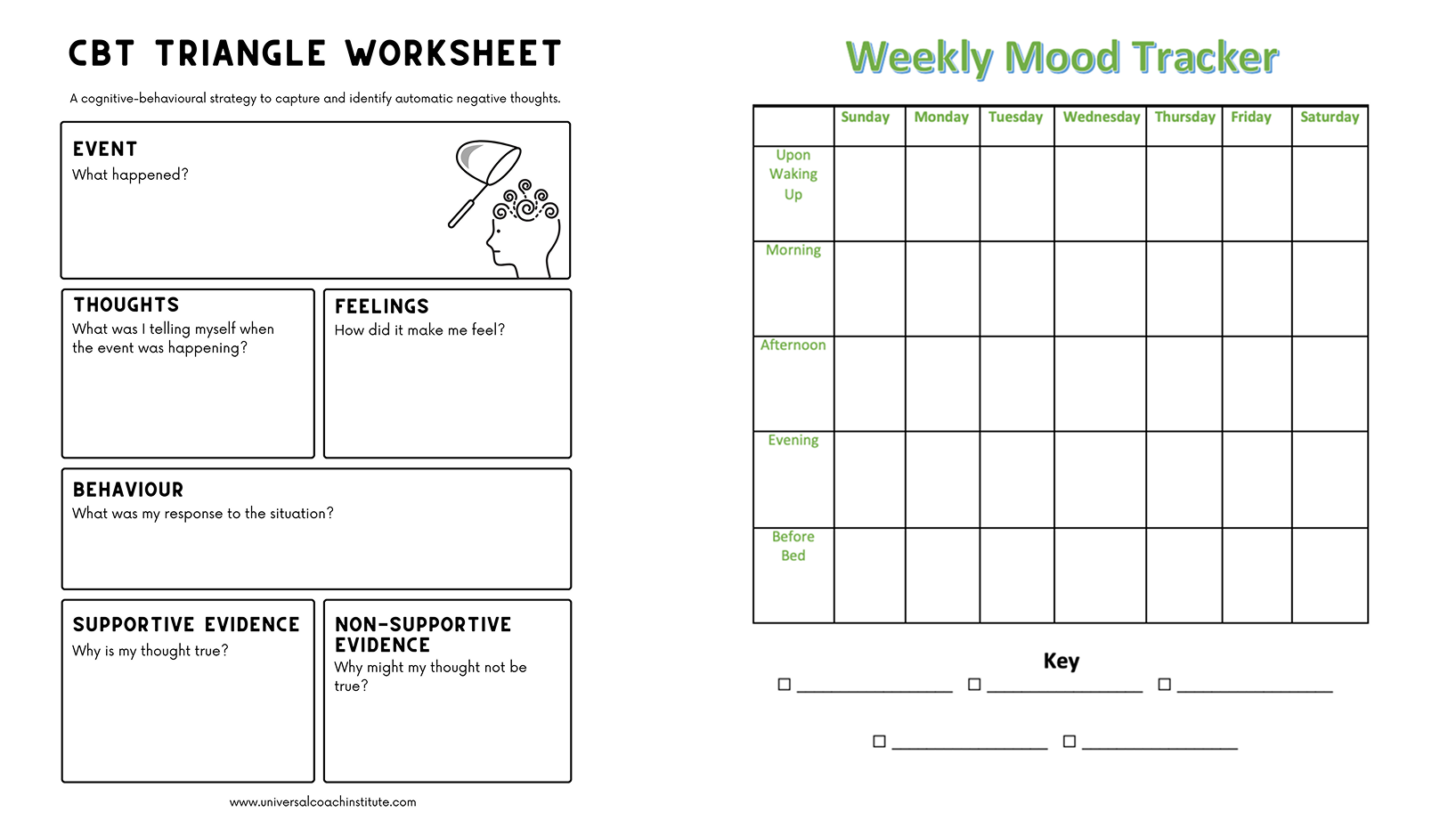}
    \caption{Examples of therapeutic homework, such as Cognitive Behavioral Therapy (CBT) worksheets and mood trackers. }
    \label{fig:homework_paper}
\end{figure*}

Therapeutic homework comprises tasks that therapists assign for clients to complete between sessions, to help them apply therapeutic concepts, build new skills, and work toward specific treatment goals~\cite{Petrik2015}. Therapeutic homework is especially central in a variety of approaches, including Cognitive Behavioral Therapy (CBT)~\cite{Freeman2007, Prasko2022}, Dialectical Behavior Therapy (DBT)~\cite{linehan2015dbt}, Solution-Focused Brief Therapy (SFBT)~\cite{Jerome2023}, Motivational Interviewing (MI)~\cite{miller2002motivational}, and Acceptance and Commitment Therapy (ACT)~\cite{hayes2006acceptance}, where clients are often asked to track behaviors, practice coping strategies, or reflect on daily experiences. Prior studies and meta-analyses have shown a clear link between homework completion and improved treatment outcomes~\cite{Mausbach2010}. Specifically, prior work has found that clients who engage more consistently with homework tend to show greater progress in managing conditions such as depression~\cite{Burns2000, Conklin2015}, anxiety~\cite{CamminNowak2013, Cooper2017}, and substance use disorders~\cite{Carroll2005}. \autoref{fig:homework_paper} shows examples of therapeutic homework, such as CBT worksheets and mood trackers. 

In practice, therapeutic homework spans several categories that differ in purpose and data needs: (1) \textit{self‑monitoring} (e.g., daily mood ratings, DBT diary cards, sleep and activity logs); (2) \textit{skills practice} (e.g., relaxation and breathing exercises, mindfulness practices, communication skills); (3) \textit{cognitive exercises} (e.g., thought records, cognitive restructuring worksheets, behavioral experiments); (4) \textit{behavioral activation and exposure} (e.g., activity scheduling, in‑vivo or imaginal exposures with SUDS ratings); and (5) \textit{psychoeducation and values work} (e.g., readings, values clarification, relapse‑prevention planning)~\cite{korotitsch1999overview, linehan2015manual, Prasko2022, bennettlevy2004oxford, martell2022behavioral, hayes2011acceptance}. 

\hl{In everyday practice, therapists manage homework in ways using a mix of paper worksheets, handwritten notes, Electronic Health Record entries, and clients' in‑session reports\mbox{~\cite{Scheel2004, Kazantzis2002, maniss2018collaborative}}. 
Studies have noted that therapists frequently struggle to track what was assigned, whether it was completed, and how clients experienced the task, especially when homework discussions must compete with other clinical priorities during limited session time, and face a lack of infrastructure that connects homework to documentation, supervision, and outcome monitoring\mbox{~\cite{Bunnell2024, Bunnell2021, Kazantzis2002, Scheel2004}}. 
In addition, many of the artifacts therapists rely on in everyday practice fall under the broader category of PGHD, such as symptom ratings, sleep records, and activity tracking. 
However, little is known about how therapists engage in sense-making around multi-format homework, and what kinds of tools could better support this process.}

\subsection{Existing Technological Approaches to Therapeutic Homework Support}

Digital tools have become increasingly integrated into mental health care to support both clients and therapists in working with therapeutic homework. On the client side, mHealth and self-help applications such as Sanvello~\cite{SanvelloApp} and Headspace~\cite{headspace} enable mood tracking, journaling, guided meditation, and psychoeducation, encouraging clients to practice therapeutic strategies between sessions~\cite{Moberg2019, ODaffer2022}. \hl{In parallel, therapy-extender applications have been developed to support clients between sessions, such as PTSD Coach, CPT Coach, and PE Coach\mbox{~\cite{Reger2013, Kuhn2014, Owen2015, Kuhn2015}}. These tools serve as early examples of blended care, where digital tools supplement in-person therapy by reinforcing skills practice and supporting continuity between appointments\mbox{~\cite{Wentzel2016, Lattie2025}}. Prior work has also examined how such tools can be used to support traditional therapeutic practices in real-world care models\mbox{~\cite{Stawarz2020}}.} On the therapist side, platforms such as SimplePractice and TherapyNotes streamline practice management tasks (e.g., scheduling, billing, documentation)~\cite{Lewis2019, MeyerKalos2024, Ridout2025} while outcome-monitoring systems such as Mirah focus on collecting standardized measures~\cite{mirah_website}. Some systems also allow therapists to distribute and retrieve digital worksheets or check-ins, enabling clients to complete assignments electronically~\cite{Lewis2019}. 

Beyond conventional apps, recent work has explored how GenAI can augment therapeutic homework~\cite{Stade2024}. Research prototypes have attempted to use LLMs to prompt reflective journaling and conversational tools~\cite{Liu2025}. For example, MindfulDiary encourages daily reflections by suggesting prompts and highlighting recurring themes~\cite{Kim2024}; MindScape combines behavioral data with generative feedback to sustain engagement with self-tracking practices~\cite{Nepal2024}. Commercial tools such as Wysa~\cite{inkster2018empathy} and Youper~\cite{youper} further embed language models in everyday conversations, offering cognitive exercises and emotional check-ins between sessions~\cite{Wang2025, Siddals2024, Chang2024, Mehta2021}. These tools aim to support clients in practicing therapeutic strategies and in maintaining continuity between appointments. Some other work has explored applying GenAI to clinical documentation, such as summarizing client visit transcripts and assisting with progress-note drafting~\cite{Kambhamettu2024, Biswas2024, Kernberg2024, Lee2024}, with ambient AI scribes reporting early efficiency gains~\cite{Tierney2024}. Early prototypes also demonstrate the potential of GenAI techniques to surface changes across records or flag risk indicators, such as mentions of suicidal ideation, which offers therapists additional cues for case formulation~\cite{Workman2023, Edgcomb2024}.

Despite these advances, most tools have primarily prioritized client‑facing engagement or focused on administrative documentation. Little work has explored how tools can be designed to support therapist‑facing monitoring of client homework and mental health counselors' perspectives and experiences with such tools. 
\hl{Prior work has begun to uncover how clinicians engage with digital tools for assessment, documentation, and treatment planning. For example, researchers have examined how therapists use digital tools to tailor between-session activities and support collaboration with clients\mbox{~\cite{Oewel2024a, Oewel2024b}} and how they appropriate technology to enable case formulation and supervision workflows\mbox{~\cite{Kuo2022}}. Other work has explored how in-home or ambient technologies can augment specific therapeutic interventions, such as behavioral activation\mbox{~\cite{Bhat2021}}, and how sensor-captured patient-generated data can facilitate clinical decision-making in PTSD treatment\mbox{~\cite{Evans2024}}. However, existing research has not yet addressed how tools can be designed to support therapists' work with clients' therapeutic homework, nor how GenAI might support these therapist-facing sense-making tasks. Our work seeks to address this gap.}

\section{Formative Design and Design Rationale of \tool}
\label{sec:tool}

Designing a GenAI-supported tool to assist therapists in reviewing client homework requires a grounded understanding of current practices, needs, and trust dynamics. Rather than treating design as a vehicle to validate predefined assumptions, we followed a design-oriented research approach~\cite{Fallman2003}, where design serves as a mode of inquiry, and the research goal is to question existing problem framings and explore new possibilities. 
\hl{We used the design process and prototype as probes to examine how therapists make sense of clients' therapeutic homework, rather than as a means to evaluate system performance alone. Such a perspective allows us to surface situated challenges, interpretive strategies, and concerns that may not emerge through conventional evaluation methods.}

Our investigation is guided by the following research questions:

\begin{itemize}[noitemsep, topsep=0pt]
\item \textbf{RQ1.} What are therapists' practices and challenges related to managing and interpreting between-session client homework data, and how should these inform the design of a homework-tracking tool?
\item \textbf{RQ2.} How do therapists perceive and interact with a GenAI-supported homework-tracking tool?
\end{itemize}

To inform the design of \tool, we conducted a two‑part formative study with practicing therapists: (i) a survey study to characterize current practices, pain points, and needs around ongoing between‑session homework; and (ii) a series of weekly co‑design sessions with three therapists in which we examined existing artifacts (e.g., paper worksheets, journal screenshots), walked through their end‑to‑end workflows for assigning and reviewing homework, and iteratively sketched/validated interface ideas. These activities grounded the design rationales and directly shaped \tool's core functionalities. All procedures were approved by our Institutional Review Board (IRB).

\subsection{Formative Study}

\subsubsection{Study Procedures}
We conducted the formative study combining survey insights with design collaboration.

\textbf{Component A — Survey.} 
We first refined an initial survey through one‑on‑one discussions with three collaborating therapists (approximately 60 minutes per session over several weeks). We then distributed the finalized survey to licensed therapists who routinely assign or review therapeutic homework. Participants were recruited through professional mailing lists, social media platforms, and direct outreach by mental health experts on our team. Participants were invited to complete an online survey via Qualtrics~\cite{qualtrics_2025} (an online survey platform). The survey included three main sections: (1) therapists' current practices in assigning and tracking homework, including typical homework types (e.g., journaling, mindfulness practices), tracking methods, and related challenges; (2) details therapists request clients to record for assigned homework, along with the rationale for why such information is considered clinically meaningful; and (3) desired technological features therapists envision would help track client progress, such as detailed completion summaries, visual representations of progress, and integration of self-reported client insights. See \autoref{sec:survey_questions} in \autoref{sec:appendices} for the detailed survey questions.

\textbf{Survey sample and use of homework:} We received 41 survey responses, of which 27 were eligible survey responses from licensed therapists who currently assign between‑session therapeutic homework.  
\hl{Specifically, 11 responses were excluded due to incompleteness. Furthermore, two participants were removed as they reported not using therapeutic homework in their practice. Finally, one response was discarded for low data quality, as manual inspection revealed the use of gibberish to bypass mandatory open-ended questions.} 

\textbf{Component B — Co‑design with therapists.}
In parallel with---and following---the survey, we held weekly 60‑minute co‑design sessions with the same three therapists throughout the design stage \hl{in person}. 
\hl{The three therapists were all licensed psychotherapists with active caseloads and 2–26 years of practice experience, spanning work in community mental health, integrated care settings, and university counseling services. Their therapeutic orientations included cognitive-behavioral therapy, trauma-informed and ACT-based approaches, and values-and-resilience-focused work. All three routinely assign and review homework in their clinical work, though the forms of homework they use vary across their therapeutic focus areas, ranging from skills practice and mood-tracking to reflective and values-based exercises.}

\begin{figure*}[h]
    \centering
    \includegraphics[width=0.95\linewidth]{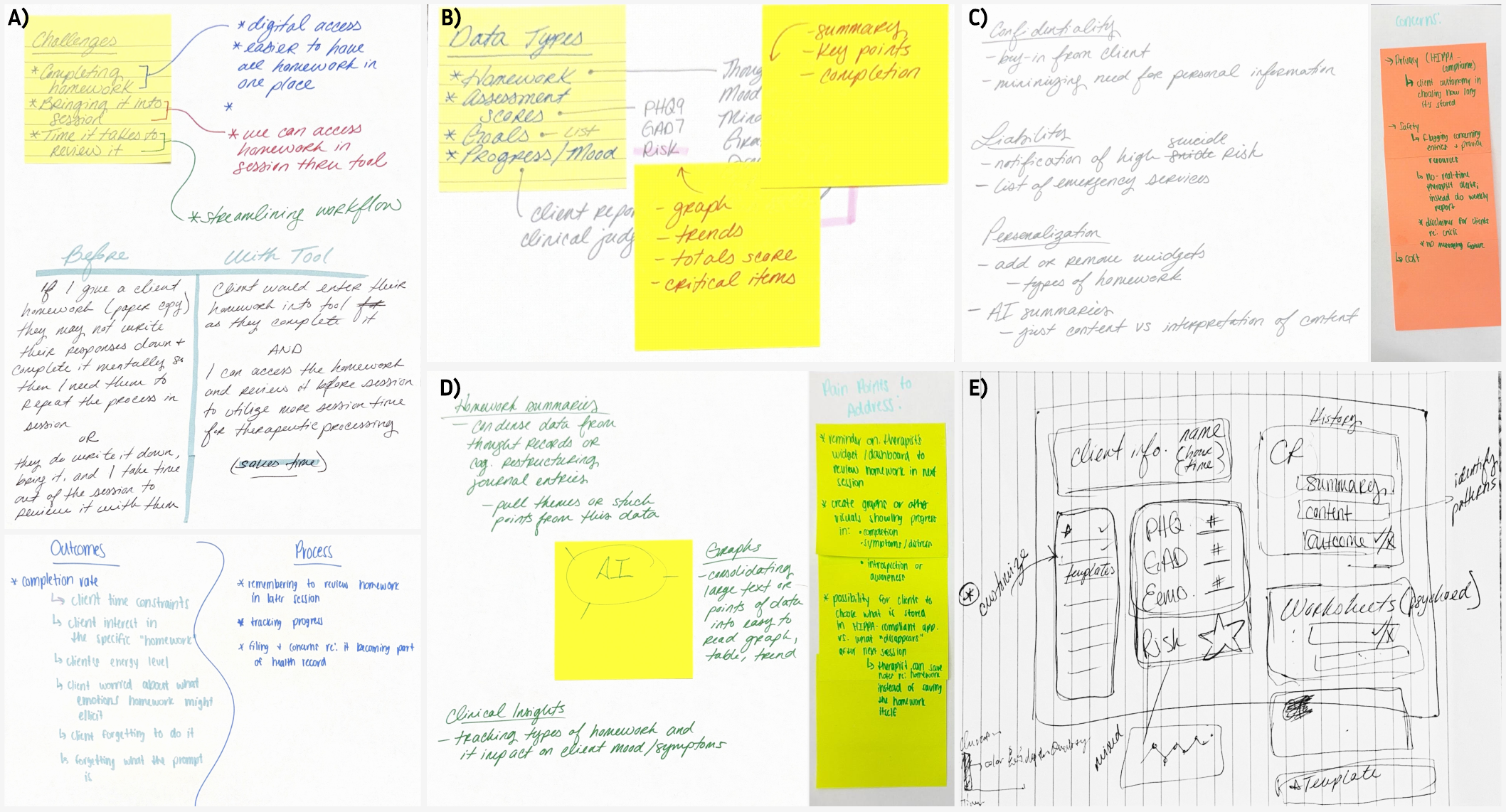}
    \caption{\hl{Examples from the co-design sessions. A) Therapists identified and discussed key challenges they encounter when reviewing and making sense of therapeutic homework in mental health counseling, as well as the types of support they would want the system to provide for each challenge. B) Therapists generated lists of data types they considered clinically valuable, discussed what information should be included in the tool, and evaluated the feasibility of capturing or representing such data. C) Therapists reflected on broader design considerations, such as information organization, interaction flow, and conditions under which the tool would or would not be appropriate for clinical use. D) Therapists specified where and how they believed AI assistance could contribute. E) Therapists drew and annotated low-fidelity sketches.}}
    \label{fig:co_design_figure}
\end{figure*}

Each session followed an artifact‑based elicitation and workflow walkthrough: therapists (i) brought examples of the homework they currently assign and receive (e.g., paper thought records, DBT diary cards, app screenshots); (ii) demonstrated how they track, review, and synthesize these materials across systems; and (iii) \hl{identified which data types would be clinically useful and sketched how the data can be presented on paper. Therapists also annotated low-fidelity sketches and discussed information hierarchy and desired workflow integration. We then translated their feedback into interface design components (e.g., what to summarize, visualize, and validate) and validated the prototype revisions in subsequent sessions. Anonymized examples of therapist-generated co-design materials are illustrated in \mbox{\autoref{fig:co_design_figure}}.}

\subsubsection{Data Analysis}

We first calculated descriptive statistics for closed-ended survey items. We conducted a thematic analysis~\cite{thomas2006general} to examine qualitative data (i.e., open-ended survey responses, co-design session notes, and annotated sketches). The lead author initially performed open coding, labeling excerpts according to recurring barriers (e.g., fragmented records, time constraints) and desired features (e.g., progress visualization, integrated views, contextual summaries). We iteratively refined these codes through regular research team discussions, reconciling discrepancies via consensus and organizing related codes into higher-level themes. We then triangulated themes across survey and co-design sources, noting whether insights converged or remained source-specific. The finalized thematic framework guided the formulation of design rationales in \autoref{subsec:drs}.

\subsubsection{Overall Results of the Formative Study}
\label{subsubsec:survey_results}

\textbf{Use of homework.} All participants indicated that homework is a routine part of treatment rather than an occasional add‑on, and most reported using multiple homework formats in parallel (median of 5 types per therapist). The most common assignments were journaling ($n=27$), mindfulness practice ($n=26$), cognitive restructuring exercises ($n=25$), and thought records ($n=21$), with fewer therapists reporting behavioral assignments such as activity scheduling ($n=14$), behavioral experiments ($n=12$), and exposure tasks ($n=7$).

\textbf{Current tracking practices.} 
When asked how they currently track homework, all therapists reported relying on client self‑report during sessions. Over half ($n=14$) also used paper worksheets, and a smaller subset (N=5) used any technology‑based tool (e.g., web or mobile apps) alongside self‑report. Several therapists noted that information was spread across paper notes, email attachments, and clients' verbal recaps, which made it difficult to keep a coherent record of what was assigned and completed over time. Together, these responses highlight that homework tracking is still largely manual and fragmented, even among therapists who regularly assign structured CBT‑style tasks.

\hl{\textbf{Details therapists ask clients to record.} 
In open‑ended responses, therapists described asking clients to record a combination of: (1) emotional states (e.g., mood or the Subjective Units of Distress Scale that measures the intensity of emotional distress or anxiety ratings before and after an exercise); (2) cognitive content (e.g., automatic thoughts, beliefs, cognitive distortions, alternative thoughts); (3) context and triggers (e.g., time of day, situation, people present, triggering events); (4) behavioral details (e.g., which skill was used, what the client did, duration, avoidance versus approach); and (5) reflections and insights (e.g., what the client learned, how an exercise affected their goals or values). Therapists emphasized that these details were primarily intended to support sense‑making.}

\hl{\textbf{Barriers to tracking and desired technology support.} 
In open‑ended responses, therapists highlighted three primary challenges: (1) \textit{Fragmentation of data}, where insights remained trapped in weekly verbal reports or scattered documents (e.g., uploaded PDFs, email threads). This challenge was particularly acute for telehealth providers, who noted the difficulty of exchanging physical worksheets or relying on clients to print materials, often leading to homework being abandoned or insights being scattered across emails and verbal reports; (2) \textit{Ambiguity of non‑compliance}, where the reliance on self‑report makes it impossible to distinguish between a client who genuinely forgot their worksheet and one who avoided the task due to anxiety or distress; and (3) \textit{Time required for review}. Therapists emphasized the importance of systematically reviewing client progress, specifically through visuals of completion rates or distress levels, to inform treatment adjustments. However, the manual effort required acts as a barrier to these insights. One therapist vividly described this as a ``time warp,'' noting that without aggregated data, it is difficult to see when a goal has stalled, risking months of stagnation without re‑evaluation. Consistent with these challenges, therapists’ interest in technology centered on tools that emphasize understanding and context. When asked to prioritize potential features, respondents overwhelmingly favored those that provide context and visibility. Frequently chosen options included tracking changes in client condition over time ($n=10$; e.g., self‑ratings of mood or self‑efficacy), access to clients' self‑reported insights ($n=10$), and visual presentations of progress data ($n=9$). Many also expressed interest in automated summaries of client progress ($n=7$) and tools that suggest possible next homework assignments based on progress ($n=10$). In contrast, fewer therapists prioritized purely quantitative metrics such as completion rates ($n=5$) or time spent on exercises ($n=4$). Therapists also noted that such features would allow them to ``chart growth,'' ``identify cognitive distortions,'' and ``make patterns easier to see'' than current methods allow. }

\hl{In the next section, we synthesize these themes, together with insights from the co‑design sessions, into three design rationales that guided the design of TheraTrack.}

\subsection{Key Insights and Design Rationales (DRs)}
\label{subsec:drs}

\subsubsection{\textbf{DR1: Centralize Heterogeneous Homework Inputs to Reduce Fragmentation with Customization}} 

\hl{As summarized in \mbox{\autoref{subsubsec:survey_results}}, therapists described homework information as scattered across various types of formats, which created inefficiencies. In both the survey and co‑design sessions, therapists asked for a single place where they could quickly find and scan relevant materials, while still tailoring which sources mattered for a given case.} 
One therapist described such difficulty, \textit{``I have to upload something into TherapyNotes and then filter through documents and open multiple tabs.''} They also emphasized that informational needs vary by orientation (i.e., the clinician's therapeutic approach, such as CBT, DBT, ACT, or psychodynamic therapy), case complexity, and session purpose; therefore, any centralized view should be \emph{configurable} rather than one‑size‑fits‑all, allowing clinicians to include only the sources that matter for a given practice or case.

\textbf{Design implication.} While self‑report remains the backbone of homework review, a centralized hub should also be able to incorporate other routinely available sources. For example, standardized measures and (when clients opt in) lightweight device summaries—to provide context without increasing therapist data entry burden.

\textbf{Translating to \tool features.} Addressing DR1, \tool consolidates heterogeneous homework inputs into one hub. Widgets support (a) homework artifacts and journals, (b) assessment timelines, and (c) optional device summaries, allowing therapists to find everything in one place without tab‑hopping. The canonical data layer behind the interface keeps raw artifacts instantly accessible (open the original entry from any summary, e.g., with \includegraphics[height=3mm]{figures/C1.pdf} supporting hover to reveal detailed notes). Furthermore, \tool enables therapists to configure the customization of sources and widgets. During onboarding, therapists can select which sources \includegraphics[height=3mm]{figures/A.pdf} and widgets \includegraphics[height=3mm]{figures/B.pdf} to include (e.g., journals, assessments, and optional device summaries), so the hub contains only clinically relevant inputs for that therapist and case. 

\subsubsection{\textbf{DR2: Make Homework Status and Progress Legible at a Glance}} 

Therapists emphasized being able to see what was assigned, what was completed, and how engagement is trending over time. They wanted a glanceable structure that can cover features like counts, timing, streaks, before/after moods for an assignment, to support preparation and in‑session check‑ins. Several highlighted that timely, visible feedback could reinforce motivation (e.g., one therapist explained \textit{``real‑time feedback could be a real boost''}), and some noted a lack of \textit{``a reliable way to monitor clients' completion or progress of assignments.''} Therapists also wanted support for monitoring long-term progress. 
One therapist acknowledged this challenge directly: \textit{``I have not yet developed a reliable way to monitor clients’ completion or progress of assignments.''}

\textbf{Design implication.} 
Interfaces should leverage visual techniques to make homework status and progress immediately legible, such as charts, timelines, and color cues. Visual summaries help therapists grasp completion and engagement patterns at a glance, while longitudinal displays (e.g., streaks, trend lines) make it possible to track trends across homework over time.

\textbf{Translating to \tool features.} 
To address DR2, we employed a clean, widget-based layout that foregrounds readability, with each widget providing a compact, glanceable view. The Homework Overview widget provides assignment‑level status (e.g., what was assigned vs. completed, recency/cadence), before/after self‑ratings where provided, and simple longitudinal trends, making progress immediately scannable and reducing the cognitive load of piecing together records. The Assessment widget encodes severity levels through color, allowing therapists to identify areas of concern at a glance. The GenAI Summary widget employs bolded keywords and text blocks to highlight salient points, ensuring therapists can quickly locate the information they need without wading through dense narratives.

\subsubsection{\textbf{DR3: Support Clinical Sense‑Making and Verification Across Modalities}}

Therapists also indicate they would like to have features to support their clinical sense-making processes. In our context, \emph{clinical sense-making} refers to the iterative process by which therapists interpret heterogeneous client inputs (e.g., journals, mood ratings, assessment scores, and health signals \footnote{Although health signals are not homework in the strict sense, therapists in our survey and co-design sessions emphasized the value of incorporating contextual indicators such as sleep, heart rate, or step counts. They noted that such signals are often directly discussed in therapy sessions to explain variations in homework engagement or outcomes.}) to (i) notice patterns across time and context, (ii) explain those patterns by forming or revising clinical hypotheses, and (iii) help decide what to do next (e.g., refine goals or assign targeted homework). In other words, therapists in our study desired functions that can help them connect what clients did with how they felt \emph{before/after} tasks and the surrounding context. They often asked clients to document their \hl{responses or experiences} before and after homework tasks. For example, one therapist described, \textit{``we explore insights in-session as to how their activity/homework was received, the effect that it had.''} Several therapists highlighted a focus on \textit{``help[ing] the client think more deeply about their maladaptive thoughts while also brainstorming, reviewing, and practicing more appropriate thought patterns''}, which is reflected in frequent mentions of terms such as ``thought'' ``helps'' and ``motivation.'' 
\hl{Therapists also expressed interest in monitoring how clients’ emotional states shifted throughout daily life, describing a need to capture the feelings surrounding specific activities and the contexts in which they occurred.}
One therapist noted the value of helping clients \textit{``track their emotional responses through the day during different events to help us understand triggering events and the actual impact of the time or situation,''} while another emphasized that knowing what clients \textit{``felt during the activity, and that tells me a lot about their progress.''} 

\hl{Consistent with the survey results in \mbox{\autoref{subsubsec:survey_results}}, therapists prioritized features that help them track changes in client condition and access clients' own insights over quantitative completion metrics, and reported relying heavily on structured reflective exercises (e.g., journaling, thought records) in their current practice.}

\textbf{Design implication.} 
\tool should go beyond displaying data (status/progress legibility) to scaffolding 
\hl{sense-making by helping therapists identify patterns, inconsistencies, and clinically meaningful relationships across inputs.} This entails transforming disparate inputs (self‑report, standardized measures, and optional device summaries) into coherent overviews that foreground clinically meaningful relationships between activities, emotions, and context; providing transparent pathways back to raw entries to enable verification; and enabling interactive, question-driven exploration aligned with clinical reasoning.

\textbf{Translating to \tool features.} Addressing DR3, in our implementation, the GenAI Summary widget provided by \tool produces source‑linked overviews (e.g., Homework Completion Trends, Journaling \& Thought Records, Assessment Patterns, optional Device Context). The GenAI Chat Assistant supports question‑driven retrieval, to help therapists answer questions like ``Has this concern come up before?'' ``Show entries where sleep dropped before low‑mood logs.'' Both features \includegraphics[height=3mm]{figures/C2.pdf} emphasize traceability---every claim links back to the originating artifacts so therapists can verify, revise, or disregard the AI's interpretation.

\subsection{Technical Implementation}
\begin{figure*}[h!]
    \centering
    \includegraphics[width=0.99\linewidth]{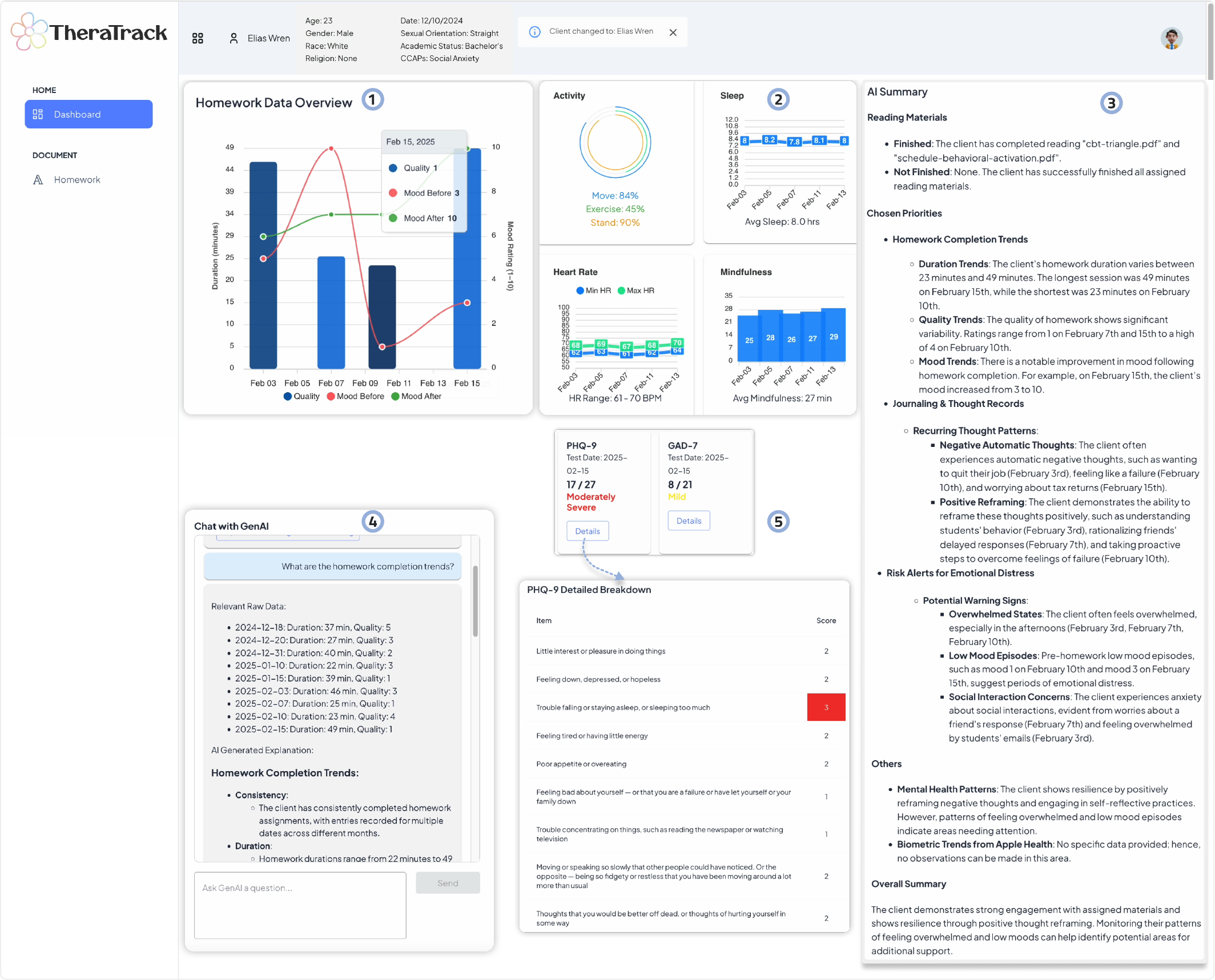}
    \caption{\hl{The TheraTrack main user interface showcases five example widgets. 1) Homework data overview: a bar chart visualizes the time required to complete each homework assignment, with color saturation indicating the client’s self-rated quality on a five-point scale. Two accompanying line graphs display the client’s mood ratings before and after completing each assignment, using a 1–10 scale in which 1 indicates a low mood and 10 indicates a very positive mood. 2) Health signals information, including activity levels, sleep patterns, heart rate, and mindfulness metrics. 3) GenAI-generated summary: information is organized according to priorities set by the therapist in their preferences; the system summarizes key points under ordered sections and consolidates minor points under an ``Others'' category. 4) Therapists can interact with a GenAI chatbot using natural language to query information about the client; responses are structured by first presenting the ``relevant raw data,'' followed by an ``AI-generated explanation.'' 5) Assessment results: when ``Details'' is selected, it expands to show all assessment items, highlighting in color any items that exceed threshold values. }}
    \label{fig:example_widgets}
\end{figure*}

\tool is a multi-page web application (see \autoref{fig:example_widgets}) which is built upon the Next.js framework~\cite{NextJS}. To unify fragmented client input across modalities, we first consolidate all user-generated data, including journaling entries, homework submissions, mood logs, and biometric summaries, into a JSON format. This JSON acts as a canonical data source, enabling downstream modules to access consistent and well-organized client information. Different data types are then selectively transformed into tailored presentation formats. For example, homework-related indicators such as task duration, engagement depth, and mood transitions are visualized through interactive visualizations in the Homework Overview widget. Raw text data from client journals or assignments remains accessible via document viewing components (e.g., embedded PDF readers), allowing therapists to inspect original submissions when needed.

Both the GenAI Summary and the GenAI Chat Assistant are powered by the GPT-4o model, accessed via Azure OpenAI Service~\cite{microsoft_azure_openai}. To support flexible and context-sensitive responses, we implemented a modular prompting pipeline inspired by LangChain~\cite{langchain}. The pipeline operates in three stages: request classification, context retrieval, and prompt construction. Incoming requests are first classified to determine intent and scope, distinguishing between GenAI Summary requests, which trigger synthesis of homework and related client data, and GenAI Chat Assistant queries, which generate context-sensitive responses to therapist questions (e.g., about historical patterns, comparative insights, or targeted analyses). Based on this classification, the system retrieves relevant homework logs, journaling text, biometric data, and therapist preferences (e.g., summary level, format, raw-data inclusion). A structured prompt is then assembled dynamically by combining 1) task-specific instructions, 2) relevant client context, and 3) the user query (for GenAI Summary, this refers to a control signal indicating whether the summary widget should be activated, rather than a free-form textual input). Prompts are issued in zero-shot mode with formatting constraints to ensure consistency. The summary module produces sectioned outputs with predefined headers (e.g., Homework Completion Trends, Journaling \& Thought Records, Biometric Patterns, Risk Alerts), while the chat assistant generates retrieval-based responses tailored to therapist queries (see Example~\ref{sec:prompt_examples}). All completions use consistent inference parameters (temperature = 0.7) to maintain stable and predictable outputs.

\section{Pilot Study}
\label{sec:Interview}

To evaluate \tool, we conducted a \hl{pilot} study with 14 mental health professionals between March and May 2025 to explore whether the designed dashboard met the practical needs and their attitudes toward, experiences with, and willingness to use \tool. 
\hl{Before recruiting therapists, we ran pilot sessions with research team clinicians to refine \mbox{\tool}'s interface flow and study procedures, such as adjusting task ordering, streamlining the tutorial, and clarifying instructions}. Below, we describe the study methods (\autoref{subsec:evaluation_study_methods}) and then present the results (\autoref{subsec:evaluation_results}). 

\subsection{Methods}
\label{subsec:evaluation_study_methods}

\subsubsection{Sampling \& Recruitment} 

We used purposive sampling to recruit practicing therapists who routinely assign and review client homework. Candidates were identified through professional mailing lists and networking platforms (e.g., LinkedIn) and via referrals from mental‑health colleagues. Each was contacted individually via direct message or email, with a brief introduction to the study and its focus on evaluating a therapist-facing tool, including the expected time commitment and the three components of the single‑session study (e.g., short tutorial, think‑aloud exploration with a simulated case, and follow‑up interview). Of the individuals contacted, 14 agreed to participate and completed the study. 

\subsubsection{Participants Overview} 
Overall, participants represented a wide range of clinical experience (1–26 years), AI familiarity, and therapeutic modalities (see participant details in \autoref{tab:pd_info}). Their exposure to AI varied from passive awareness to exploratory or clinical use. Most therapists reported using multiple cognitive-behavioral exercises in practice, with Cognitive Restructuring, Mindfulness Exercises, and Journaling \& Thought Records appearing most frequently. Therapeutic homework assignments similarly spanned a diverse set of tasks, including mood tracking, gratitude journals, behavioral experiments, and breathing exercises.

\begin{table*}[h!]
\centering
\footnotesize
\caption{Background information for the 14 therapists who participated in our study. Each row represents a unique participant. The table includes: (1) participant ID with gender (F: female, M: male); (2) years in practice, indicating total clinical experience; (3) AI experience, summarizing prior exposure to or use of AI tools; (4) therapeutic focus areas, describing the modalities each participant emphasizes in their work; and (5) example therapeutic homework (i.e., typical client activities assigned between sessions). Focus areas and homework types were selected from a predefined list in the onboarding survey, where participants could choose multiple options relevant to their practice (see \includegraphics[height=3mm]{figures/A1.pdf} in \autoref{fig:overview_of_dashboard})}.

\rowcolors{2}{gray!10}{white}
\begin{tabularx}{\textwidth}{l c l L L}
\toprule
\textbf{Participant} & \textbf{Years in Practice} & \textbf{AI Experience}$^\star$ & \textbf{Therapeutic Focus Areas} & \textbf{Example Therapeutic Homework} \\
\midrule
P1 (F) & 2 years & Heard only &
Cognitive Restructuring, Mindfulness Exercises, Behavioral Activation &
Thought Records, Mood Tracking, Relaxation \& Breathing Exercises \\

P2 (F) & 5 years & Exploratory Use &
Cognitive Restructuring, Mindfulness Exercises, Journaling \& Thought Records &
Thought Records, Gratitude Journals, Mood Tracking, Relaxation \& Breathing Exercises \\

P3 (F) & 18 years & Used non-clinically &
Mindfulness Exercises, Behavioral Activation, Journaling \& Thought Records and Other &
Thought Records, Behavioral Experiments, Gratitude Journals, Mood Tracking, Relaxation \& Breathing Exercises and Other \\

P4 (F) & 1 year & Used non-clinically &
Journaling \& Thought Records &
Thought Records, Gratitude Journals, Mood Tracking, Relaxation \& Breathing Exercises \\

P5 (F) & 4 years & Exploratory Use &
Cognitive Restructuring, Mindfulness Exercises, Behavioral Activation, Journaling \& Thought Records and Other &
Thought Records, Behavioral Experiments, Gratitude Journals, Mood Tracking, Relaxation \& Breathing Exercises \\

P6 (M) & 26 years & Heard only &
Mindfulness Exercises, Exposure Therapy, Journaling \& Thought Records and Other &
Thought Records, Behavioral Experiments, Mood Tracking \\

P7 (F) & 1 year & Heard only &
Cognitive Restructuring, Mindfulness Exercises, Journaling \& Thought Records &
Thought Records, Mood Tracking \\

P8 (F) & 11 years & Used non-clinically &
Cognitive Restructuring, Mindfulness Exercises, Journaling \& Thought Records &
Thought Records, Gratitude Journals, Mood Tracking, Relaxation \& Breathing Exercises and Other \\

P9 (F) & 2 years & Used clinically &
Mindfulness Exercises, Journaling \& Thought Records and Other &
Thought Records, Mood Tracking, Relaxation \& Breathing Exercises \\

P10 (F) & 5 years & Used non-clinically &
Cognitive Restructuring, Mindfulness Exercises, Exposure Therapy, Journaling \& Thought Records and Other &
Behavioral Experiments, Mood Tracking, Relaxation \& Breathing Exercises \\

P11 (M) & 10 years & Heard only &
Cognitive Restructuring, Journaling \& Thought Records &
Thought Records, Gratitude Journals \\

P12 (F) & 15 years & Used non-clinically &
Cognitive Restructuring, Mindfulness Exercises, Journaling \& Thought Records and Other &
Gratitude Journals, Mood Tracking and Other \\

P13 (F) & 8 years & Used non-clinically &
Mindfulness Exercises, Journaling \& Thought Records &
Thought Records, Gratitude Journals, Mood Tracking, Relaxation \& Breathing Exercises and Other \\

P14 (F) & 1 year & Heard only &
Cognitive Restructuring, Mindfulness Exercises, Journaling \& Thought Records &
Thought Records, Behavioral Experiments, Gratitude Journals, Mood Tracking, Relaxation \& Breathing Exercises and Other \\
\bottomrule
\end{tabularx}

\vspace{1ex}
\begin{minipage}{\textwidth}
\footnotesize
$^\star$\textit{Heard only}: Aware of AI tools through others or media, but never used in any context. 
\textit{Used non-clinically}: Applied for general tasks (e.g., summarization, brainstorming). 
\textit{Exploratory Use}: Self-driven experiments in mental health settings, not client-facing. 
\textit{Used clinically}: Integrated AI into therapy or client-related work.\\
\end{minipage}
\label{tab:pd_info}
\end{table*}

\subsubsection{Study Procedures}
\textbf{Step 1. Pre-study Tutorial ($\sim$3-5 min).} 
We introduced participants to \tool's purpose and structure, walking them through key sections: the onboarding process (survey and widget selection) and interactive exploration. We demonstrated core functionalities, such as viewing client homework, reviewing behavioral data, accessing GenAI summaries, and using the chatbot, \hl{and clarifying how to interpret key interface elements (e.g., the meaning of color cues and numerical mood values).} This session was designed to ensure a clear understanding of \tool's intended use and its potential workflow integration. Participants were encouraged to ask questions at any point, to clarify their understanding before exploring \tool on their own. 

\textbf{Step 2. Participant Interacting with \tool ($\sim$15-20 mins) using Think-aloud Method.} 
After the tutorial, participants engaged in a self-guided exploration of the dashboard. They utilized a simulated client case, complete with example data and submissions drawn from our prior research, and were instructed to interact with the system as if in a real-world scenario. This involved navigating all sections, reviewing the provided information, and evaluating the dashboard's potential fit within their existing therapeutic practices. Participants were specifically asked to consider the relevance and utility of different sections, identify any areas of confusion or misalignment with their needs, and suggest any potentially missing features or workflows.

\textbf{Step 3. Interview and End Survey ($\sim$45-50 min).} 
Following the think-aloud, we conducted a semi-structured interview to capture further insights to further examine participants’ perceptions of \tool. Participants were asked to reflect on \tool's alignment with their therapeutic goals, their confidence in using it in practice, their attitudes and trust in such tools, and any recommendations for design improvements. 

To complement the qualitative data, participants then completed a five-point Likert battery adapted from the Health Information Technology Usability Evaluation Scale (Health-ITUES)~\cite{Schnall2018}, as well as items assessing trust in AI-generated outputs given evidence that trust influences perceived usefulness and technology use in mental health settings~\cite{WangY2025}. See \autoref{sec:likert} in \autoref{sec:appendices} for specific measurements. At the end of the study, participants were compensated with a \$25 gift card as a token of appreciation. 

\subsubsection{Data Analysis} 
All interviews were audio-recorded and transcribed for subsequent analysis. Similar to our analysis of the formative survey study, we conducted a thematic analysis~\cite{thomas2006general} of verbatim transcripts, focusing on therapists' experiences with \tool. \hl{First, the lead author carefully read the transcripts multiple times to become familiar with the data to capture initial impressions of \mbox{\tool} among participants.} 
\hl{Then the lead author performed inductive, line-by-line coding to identify key themes. For each segment, we applied descriptive codes that captured (i) the feature or workflow being discussed (e.g., homework overview, GenAI summary, various widgets) and (ii) the type of response (e.g., perceived usefulness, trust, concerns or tensions, suggestion for change, or appropriation pattern). Next, related codes were clustered into higher‑level thematic categories, such as perceived usefulness and cognitive off‑loading, trust and verification practices around AI‑generated insights, tailoring and appropriation patterns, and privacy and implementation concerns. These categories were informed by, but not constrained to, the themes from the formative survey and co‑design work, allowing us to trace how previously identified needs (e.g., fragmentation of homework data, cognitive burden of review, desired technology support) played out when therapists interacted with the working prototype. The evolving codebook and candidate themes were reviewed in regular analysis meetings with the research team. In these meetings, we compared interpretations, merged overlapping categories, and refined theme boundaries; disagreements were resolved through discussion with reference to the original transcript context.}

\subsection{Pilot Study Results}
\label{subsec:evaluation_results}

Overall, therapists found \tool highly usable, visually intuitive, and valuable, showing strong enthusiasm for future use, with minor reservations reflecting individual contexts. In terms of overall utility, therapists found \tool highly navigable ($n=11$, 79\%) and 64\% of participants ($n=9$) strongly agreed that the graphs and metrics were visually clear. Regarding perceived usefulness, 79\% of participants ($n=11$) strongly agreed that the GenAI summaries helped reduce their workload, and 71\% ($n=10$) found the system efficient for locating information. 

Most participants ($n=9$, 64\%) gave the highest rating to the \tool's ability to support homework tracking. All participants expressed neutral to positive views on user control, \hl{with several noting that \mbox{\tool} felt easy to operate and that they could manage its functions in ways that suited their own preferences}. When asked about the likelihood of long-term integration, half of the participants($n=7$, 50\%) showed strong enthusiasm and a clear willingness to use the system. However, three participants were more reserved, reflecting possible uncertainties or context-specific limitations, \hl{such as questions about how well it would fit with their existing practices and uncertainties about client receptivity}. A similar pattern emerged regarding peer recommendation, with a majority ($n=9$, 64\%) strongly agreeing they would recommend \tool to others. 

Our interviews with therapists further revealed three key findings regarding their experiences with \tool. First, by reducing the burden of manually filtering information and synthesizing client homework data, therapists felt more prepared for sessions and could better focus on proactive, forward-looking planning. Second, although therapists consistently reported trust in the AI-generated outputs, their reasons for trusting the tool varied based on personal expectations, comfort with technology, and verification practices. Third, participants emphasized the importance of customization. They noted that the value of \tool depended on how flexibly it could fit different clinical contexts, such as private session preparation versus client-facing discussions, as well as on therapist experience levels and case-specific needs.

\subsubsection{Perceived Usefulness \& Cognitive Off-loading}

Participants described \tool's value as an integrated system that streamlined their clinical workflows. They particularly emphasized its usefulness in addressing some of their most pressing and cognitively demanding tasks, such as filtering, synthesizing, and interpreting heterogeneous homework data efficiently. 

\textbf{\hl{Potential to} Alleviating Cognitive Burden by Automating Laborious Work.}
Therapists highlighted the substantial cognitive demands of manually integrating fragmented client data, which required navigating multiple documents and relying heavily on memory. They explained that constant manual effort leads to a \textit{unstructured workflow}, where therapists could only view one document at a time rather than the client’s whole story. As P1 noted, \textit{``Because when I am going [through information], I have to go through things one at a time.''} 
\hl{Therapists described feeling that the memory-dependent process made it harder to connect the dots to identify emerging patterns.}

\hl{Our participants felt that \mbox{\tool} had the potential to address these inefficiencies by aggregating client information into an accessible platform.}
\hl{Some therapists expressed concern that, in their current workflow, especially in the context of busy caseloads, they often felt unable to devote enough attention to every document or note, and important insights may be overlooked while less critical details sometimes receive disproportionate attention.} As P1 noted, manual review required \textit{``going through things one at a time,''} whereas \tool provided \textit{``an idea of the overall pattern or something that I am interested in''}, eliminating the need to \textit{``try to [piece everything together] to get to this point.''}

\hl{Our participants described how the visualization widgets and GenAI-generated summaries helped collate a client's homework status into an organized and digestible overview.}
Rather than manually reconstructing a narrative across multiple homework assignments, therapists are presented with a synthesized snapshot that highlights progress, struggles, and emerging themes that require attention. 
For example, P9 explained, 
\begin{quote}
    \textit{``I think with the AI summary, it is helpful to track growth and also see where the stuck points are. [For example, in the] last six months, they have been saying the same automatic thought. I will say, `Let's dig a little deeper as to why that thought is still there.' So, I think that is the advantage I am seeing with \tool, especially being able to see the homework visualization and also get summaries. It helps you and your clients stay on the same page. (P9)'' }
\end{quote}

Moreover, when it comes to long-term data tracking, traditionally, therapists face various challenges, such as inconsistent record-keeping and difficulty in consolidating data from multiple sessions. However, \hl{therapists perceived \mbox{\tool} as enabling a more centralized form of tracking of data, }making it easier to monitor progress and identify patterns over time. For example, if a therapist forgets a detail or feels unsure about something in a client’s history, they can directly ask the GenAI chatbot, and receive immediate and targeted information, which decreases the need to sift through multiple notes or files. As P14 described, \textit{``by asking the AI chat assistant [in a natural conversation], this could present as a refresher, keep in mind what the initial reason is.''}
P14's comment suggests that she appreciated how easily accessible client information provided by the AI chat could serve as a quick reference or memory aid, helping them reorient to the client's core therapeutic goals or initial presenting concerns. 
\hl{Participants described this as helping them redirect attention towards} concentrating on higher-level clinical considerations rather than on administrative retrieval tasks.

\hl{At the same time, therapists in our study described a different kind of cognitive load that accompanied the use of AI-generated summaries. Rather than accepting outputs at face value, several engaged in cross-checking, comparing AI interpretation against their own reading of the client's homework or recollections of past sessions. As P5 mentioned,}

\begin{quote}
    \hl{\textit{``I just wanted to make sure I am looking at things right [and] did not miss [anything], especially with homework completion trends.'' (P5)}}
\end{quote}

\hl{Therapists like P5 noted that when the summary emphasized themes that diverged from their impressions, they became more cautious in interpreting the output. They described the need to pause and think thoroughly about why the framing differed, without assuming that either interpretation was necessarily more accurate. As P7 described, \textit{``[I am taking what] the client says and measuring [it] up to this AI summary, seeing which patterns are similar [and] which patterns are not. But that does not mean that this AI summary is wrong.''} In these cases, the summaries introduced additional interpretive work, requiring therapists to reconcile multiple perspectives without assuming that either one should dominate.}

\textbf{Session‑Ready Summaries that Shift Therapy from Review to Proactive Planning.} 

\hl{Our participants perceived that the cognitive effort saved through \mbox{\tool} could support a more proactive workflow, shifting their approach from merely reactive data review to active synthesis and clinical planning. Therapists described this perceived usefulness as particularly noticeable during pre-session preparation.}
Pre-session preparation refers to the process therapists go through before meeting with their clients~\cite{arttherapyresources2023session}. During this time, therapists review and synthesize relevant client information, such as homework completion, mood tracking data, journal entries, or health records, to form a comprehensive understanding of the client's current state, recent progress, and potential challenges to inform the upcoming session. 

In our study, rather than examining isolated data points, therapists described actively synthesizing information across different widgets to form a cohesive understanding of their client's status. For example, therapists first reviewed visual summaries (e.g., homework completion, mood trends), then moved to GenAI-generated summaries to understand underlying reasons or context. Finally, therapists might cross-reference these insights with health data to identify correlations between physical symptoms and observed behaviors. P13 explained,

\begin{quote}
    \textit{``I find \tool beneficial because I am a very visual person [who] likes having the charts. The graphs give me a brief overview and [help] shape the lens through which I read the AI summary as well. Then I get a little bit more detail in the AI summary. [Because of this,] I have a better idea of what I want to skim and look for in the AI summary. [For example,] if I saw that the client’s sleep was lower than usual, I know to look in the AI summary for what factors might be influencing their sleep. (P13)''}
\end{quote}

\hl{Several participants felt that having synthesized information before sessions influenced how they approached in-session discussions, where synthesized understanding was described as equipping the therapists with potent talking points for their next session.}
By having greater context into a client's struggles, therapists could more readily shift the therapeutic dialogue. As several participants explained, this enabled them to bypass the time-consuming phase of simply figuring out ``what is going on.'' 
As P6 explained, \textit{``the ability to track data allowed [them] to maximally, optimally use the therapeutic session with them in the most relevant way.''}
Instead, they can begin the session with targeted, insightful questions that cut directly to the core issues, focusing the precious session time on exploring ``why this happened'' and determining ``what we should do next.'' 
P6 further emphasized that accessing \textit{``great data before moving into that therapeutic session''} allows them to \textit{``make the most out of the 50-minute session.''} They noted that this preparation \textit{``enhances the richness and the effectiveness of staying focused on therapeutic progress and the therapeutic goal, [and] helps the therapist stay more attuned to the client,''} which they described as \textit{``always beneficial to the client.''}

\hl{Participants felt that these insights could help them use session time more purposefully. }Rather than spending initial session minutes recalling previous details or trying to piece together a client's narrative on the spot, therapists can immediately focus on critical therapeutic issues, deeper reflection, and interventions. 

\hl{However, participants also described moments when the insights provided by the \mbox{\tool} presented challenges to their therapeutic processes. While summarizing and visualization helped therapists notice recurring patterns or bottlenecks, therapists emphasized that clients may come with immediate questions that do not quite fit the data-driven therapeutic agenda. Several therapists were concerned about how to strike a balance between responding to what clients wanted to talk about in the moment while also addressing recurring issues in homework data. As P7 said, }

\begin{quote}
    \hl{\textit{``I think balancing those two spaces is where would be interesting. I know that you did that homework last week and it said this, and these are all important things, but you are in here today with me because you want to talk about this new thing that happens perhaps even right before the session and that is most immediate and what you want to talk.'' (P7)}}
\end{quote}

\hl{Therapists described treating these insights cautiously, selectively incorporating them into therapy and supporting rather than overshadowing the client's immediate priorities. As P11 said, \textit{``Like if they do not have a lot that they are kind of talking about. And then of course, one of the things I am gonna do is use this graph, especially where sheets went up.''}}

\subsubsection{Trust \& Verification Practices}
Eight (57\%) therapists in our study reported strong trust in AI-generated summaries, while others indicated moderate confidence. Participants attributed their trust largely to the perceived low-risk nature of \tool's functions. When contrasted with higher-stakes AI applications in mental health, such as therapeutic chatbots interacting directly with clients, \tool was considered lower risk because it functioned as a supportive aid for therapists. \tool's primary role was framed as administrative and summative, such as processing client homework and structuring data. The therapist remains the sole decision-maker and intermediary, using the AI's output as a draft or a starting point. For example, P5 said, 
\begin{quote}
    \textit{``I am not just blindly trusting what the AI says, `Oh, AI says this, so that's how it is.' [Instead,] I look at the detailed logs of thought records [in AI summaries], which is great because it shows me exactly what parts clients are engaging with and when. [Then,] I can actually see their progress and how their thinking changes over time. (P5)''} 
\end{quote}
The design keeps the locus of control and clinical responsibility in the hands of the human expert, which in turn lowers the perceived risk and mitigates anxieties about potential AI errors impacting client care, as P11 echoed, \textit{``It just gives you the summary of their homework. It is not trying to replace the therapist.''}

Participants described two pathways toward building trust in the AI. Those initially skeptical actively built trust through deliberate verification. They cross-checked GenAI-generated summaries against visualization widgets or directly examined the raw client data through the GenAI assistant to ensure accuracy. P1 explained this verification process clearly:
\begin{quote}
    \textit{``Sometimes I worry about the assumptions the AI is making or the connections it draws. Having the ability to ask, `What information did you use to make that assumption or connection?' [allows me to decide] if I agree or disagree, based on my own clinical reasoning. It does really help build my trust. (P1)''}. 
\end{quote}

Conversely, for participants who were more readily inclined to trust the AI, confidence was established by the system's credible presentation and then reinforced by its underlying transparency. As P3 explained, 
\begin{quote}
    \textit{``I did not have a reaction to not trusting the summary. The first impression is very good with the AI summary for me. (P3)''} 
\end{quote}
For participants like P3, the detailed GenAI summaries provided an impression of reliability. Then, initial trust was solidified by the \tool's design, which grounded the statement in accessible raw data. Participants found it difficult to imagine \tool could hallucinate because the direct link to the client's data allowed them to understand the ``why'' behind any AI statement. For example, P13 pointed out, \textit{``Since [\tool] gives me the summary and supports it with raw data, I trust it.''} Even for these trusting participants, the ability to verify was important. If they ever felt the AI made an unwarranted assumption, they could bypass the interpretation and review the raw data themselves. P12 explained, 
\begin{quote}
    \textit{``Well, I think it is because the AI bases its feedback on the data provided by the client, as well as other relevant inputs and models. Knowing where the data comes from and seeing how well the AI performs helps build my trust. (P12)''}
\end{quote}
These findings suggest that \tool's emphasis on transparency, clinician oversight, and easy data verification played key roles in fostering therapists' trust and confidence in the system.

\subsubsection{Tailoring \& Appropriation Patterns}
Rather than a one-size-fits-all approach, participants described distinct patterns of tailoring and appropriation that emerged as they integrated \tool into their workflows. Both perceived value and desired functionality of \tool varied not only between private analytical work and client-facing interactions, but also across different stages of therapy and levels of clinical experience. 

\textbf{Private Analysis vs. In-Session Communication.} Therapists described two potential ways of using \tool: privately during session preparation and collaboratively during client sessions. During private preparation, therapists preferred a detailed, data-rich environment, including extensive GenAI-generated summaries, longitudinal progress charts, and comprehensive homework data. 
\hl{Participants felt that having access to more information helped them engage in deeper clinical reflection. For example, P12 said, \textit{``This is definitely more comprehensive, more precise, [and] all in one place, versus me looking at my notes and then [having the] assessment somewhere else. It might give me some pointers to address, so then that helps the client go deeper.''}}
During client sessions, therapists opted for selective visibility, sharing only relevant visualizations or progress highlights to facilitate meaningful conversations without overwhelming clients. As P1 articulated, 
\begin{quote}
    \textit{``I would probably show [my clients] something without all the other options visible. I would not want them reading the AI summaries. But I would want to show them [specific things]---especially if I am trying to make a point. If a certain type of homework seems really helpful, I like to point that out. Or, if their sleep went way down and their mood was much lower, I would show them the relationship between sleep and mood and say, `This seems like something we should work on.' That is another reason I like to be able to manage my widgets, so I can choose what to share or show, without having everything else on the screen. (P1)''} 
\end{quote}
The interface shifts from a tool of analysis to a tool of communication. \hl{For example, P9 said, \textit{``I think that could definitely be interesting to present to new people who [might] just not know why biometric data is important.''} And P13 echoed, \textit{``I would want to talk to them about what is hindering their sleep [and] what they are [experiencing], so that I know if we need to provide some psychoeducation around sleep hygiene.''}} Displaying dense text or complex metrics during a session can be counterproductive as it risks making the client feel like a specimen being analyzed rather than a partner in their own care. \hl{As P3 figured, \textit{``But I am always thinking, if a client wants to see something like this? How does that get perceived? And could it potentially be harmful?''}} This can disrupt the therapeutic alliance and impede the natural flow of conversation.

\textbf{Appropriation by Clinical Experience Level.}
We found that therapists' levels of clinical experience appeared to be associated with how they interacted with \tool and GenAI-generated insights. Seasoned therapists (i.e., participants with more than three years of practice), with well-developed intuition and established workflows, primarily valued the tool for its efficiency and cognitive offloading. For them, quickly surfacing key information allowed more mental space for higher-level clinical thinking and strengthened client engagement. P6 explained, 
\begin{quote}
    \textit{``The goal is to get to the essential data faster to maximize time for higher-order thinking and client interaction. (P6)''}
\end{quote}

Novice therapists, in contrast, viewed the GenAI summaries as helpful validation and an additional perspective. For these less experienced clinicians, the AI offered reassurance, helped double-check their interpretations, and increased confidence in their therapeutic decisions. As P7 noted, 
\begin{quote}
    \textit{``Sometimes the AI brings up something I had not considered. It helps me double-check my own thinking and gives me more confidence in my plans. (P7)''}
\end{quote}

\hl{However, therapists also raised concerns that over-reliance on AI-generated interpretations might be particularly challenging for novice practitioners. They described how newer clinicians, who are still developing their own analytic frameworks, may feel inclined to defer to the AI's presentation of client issues. As P7 further explained, \textit{``I was thinking about the new clinicians, because I just had an experience [with] one of our very new master's students who [thinks] whatever the book says is the truth. [That] is where they get really stuck---[thinking] `the information I found out says it must be true,' without giving room for their client's experience.''} Therapists worried that in similar ways, a neatly organized AI summary might inadvertently reinforce premature conclusions or narrow interpretations if novice clinicians lean on it without sufficiently engaging with the client’s lived experience.}

\subsubsection{Challenges in Customization}

\textbf{Privacy as a Persistent Concern.}
While participants generally appreciated \tool's flexibility and the ability to tailor interactions, this customization raised critical privacy concerns. For example, P4 valued having a chatbot \textit{``private and just for me''} as a way to \textit{``eliminate the fear of AI,''} yet also noted that greater customization raised concerns about how client data would be handled.

Our participants emphasized that behind-the-scenes automated processing must be approached with care, in terms of data storage, visibility, and security. As P3 highlighted: 

\begin{quote}
\textit{``This is a therapy record. So anything that gets recorded here needs to be treated with the same level of care and protection.'' (P3)}
\end{quote}
In this light, customization became a double-edged sword: while greater configurability allowed therapists to shape the system to their needs, it also raised difficult questions about what information is surfaced, where it is stored, and who ultimately has access to it. P2, for example, cautioned that ethical restrictions remain non-negotiable, especially for high-risk clients:

\begin{quote}
\textit{``Sending messages to clients is important, but that comes with boundaries. With certain high-risk or severe cases, we are very restricted, both ethically and legally, on what can be shared and how it is stored.'' (P2)}
\end{quote}

\hl{Therapists in our study also emphasized that data should originate from the client rather than be inferred or repurposed, reinforcing a norm that clients maintain agency over how their information is represented. As P1 said, \textit{``They know [what] information they are putting in, and then they [also] have the autonomy of deciding how much information they want to give or not give to the program. [And] I am sure that they will also see the user agreements about their privacy.''}}

In short, therapists stressed that more information could be valuable only if it was carefully contained. As P5 summarized, they wanted the system to \textit{``have more information''} but at the same time ensure it was \textit{``not spreading the information.''}

\textbf{Supporting Therapists’ Evolving Needs.} While therapists found the onboarding interface straightforward, several raised concerns about its rigidity, feeling it did not adequately capture the diversity of their clinical orientations. Therapists desired greater flexibility in specifying their practices beyond predefined categories. For example, P12 explained:
\begin{quote}
    \textit{``Where it says `Other,' I would like to actually specify `Other.' To be specific, because my approach---I do not see it there. I do IFS (Internal Family Systems), for example, that is not there. A lot of therapists might not be only using cognitive restructuring or exposure therapy. But there [are] many more: motivational interviewing, solution-focused...So you might need more on that list.'' (P12)}
\end{quote}
P8 echoed a similar sentiment, noting the need for additional assessments specific to different therapeutic modalities:

\begin{quote}
\textit{``These [on-boarding survey items] are right on point, but I do have others I use that are not on here, like the PCL (PTSD Checklist) for trauma or OCI-R (Obsessive-Compulsive Inventory-Revised) for OCD. I think this is interesting.'' (P8)}
\end{quote}

Beyond these specific needs, our participants mentioned that they consider their interaction with \tool might also evolve throughout the therapeutic relationship. Here, the therapeutic relationship refers to the collaborative alliance between therapist and client, evolving dynamically as treatment progresses~\cite{Bordin1979}. Clients' informational needs vary across different stages. In the initial phase, for example, therapists often focus on case formulation and building rapport. During this period, a detailed GenAI summary that unpacks its reasoning can be invaluable for identifying core issues. As therapy progresses into the intermediate stages, the focus might shift to tracking the application of specific skills or monitoring adherence to a behavioral plan. The \tool's utility then hinges on its ability to visualize targeted metrics. For example, P14 said, 
\begin{quote}
    \textit{``In the beginning, I needed the full, detailed AI summary to really trust what the system was telling me [about the client]. When I have used it for a while, I would prefer a `highlights' version to get a quicker sense of things. (P14)''} 
\end{quote}
This temporal dimension is fluid even within a single case, requiring the dashboard to be flexible enough to move between detailed and summary views on demand, as P14 continued, \textit{``My needs might even change back if I get a particularly complex case where I want the AI to show me all its reasoning.''}
These findings shed light on the importance of providing a more nuanced onboarding process and adaptive interfaces that can flexibly evolve in response to therapists' changing clinical requirements over time.

\section{Discussion}
Our findings demonstrate the utility of \tool in helping therapists efficiently review, organize, and interpret fragmented therapeutic homework data. Beyond demonstrating immediate benefits for clinical preparation and workflow, the findings also offer broader insights into how GenAI-powered tools can be designed to support therapist-facing interpretation without undermining core therapeutic principles. While AI can assist therapists in navigating complex, multi-format client inputs, its integration must be carefully calibrated to preserve the interpretive authority of clinicians, support relational dynamics in therapy, and uphold ethical standards of transparency, consent, and accountability. Based on these insights, we discuss design implications for therapist-centered and auditable AI, patterns of appropriation shaped by clinical context and experience, and ethical considerations for maintaining oversight, protecting client trust, and avoiding unintended harms.

\subsection{Design implications for therapist‑centered and auditable AI}

Our findings highlighted several opportunities for supporting therapist sensemaking and workflow efficiency through potential AI integration. Participants reported that shifting from manually piecing together fragmented artifacts to quickly interpreting integrated summaries and visuals enabled proactive therapeutic planning. This finding aligns with prior HCI and personal informatics literature emphasizing progressive disclosure---providing an overview first, followed by details-on-demand, to support sensemaking and reduce cognitive load~\cite{shneiderman2003eyes, zhang2018idmvis}. The design implication here is to reinforce a distinct shift from reactive review (e.g., scanning and understanding disparate artifacts during sessions) toward a pre-session ``briefing'' model, providing summaries, glanceable trend visuals, and rapid access to source entries. At the same time, participants also cautioned that while AI-generated summaries enhanced pre-session efficiency, they should not constrain the interpretive space during sessions. In many therapeutic modalities, collaborative homework review is not merely a progress check but a joint meaning-making practice~\cite{Kazantzis2018, Westra2007}. To support collaborative rhythms, future systems could offer adjustable formats that allow therapists to choose between concise overviews for private preparation and open-ended prompts that invite client dialogue during sessions. For example, offering a simplified ``conversation mode'' for client-facing interactions can maintain therapeutic alliance and reduce client overwhelm while ensuring therapists remain in control of clinical interpretation.

\textbf{Calibrate trust through traceability and question‑driven verification.}  
Another key finding was therapists' emphasis on calibrated trust through traceable AI outputs. While participants trusted the AI-generated summaries, their trust was highly dependent on verification capabilities---namely, direct links from AI assertions back to original source data. In fact, these verification practices reflect the emerging design principle of  ``explanatory debugging'' in HCI literature, where AI systems expose not only their outputs but also their rationale, helping users to actively interrogate and refine their mental models of the AI~\cite{kulesza2015principles, liao2020questioning}. Designers should thus prioritize provenance by default, offering persistent inline anchors to original client inputs, as well as supporting interactive queries (e.g., ``Where did this come from?'' and ``Show me evidence''). By designing explicitly for traceability and verification, future tools can foster appropriate reliance, reduce automation complacency, and maintain clinical accountability.

\hl{However, our findings suggest that therapists' confidence in AI summaries was often better characterized as trust in their own oversight than trust in the model itself. Participants described treating the summary as a starting point and relying on traceability to audit, confirm, or reject claims based on their clinical judgment. In this sense, traceability supports calibrated reliance by making verification feasible, but it does not guarantee correctness. Prior work shows that, even with provenance links, a system can still omit relevant context, overweight salient snippets, or produce plausible but inaccurate interpretations; links may simply help users detect errors when they choose (and have time) to check~\mbox{~\cite{Asgari2025, Topaz2025}}. Because our evaluation used a simulated case with populated inputs, participants may not have encountered ambiguous, sparse, or inconsistent data situations as in real-life practice, which could have shaped perceptions of completeness. Future implementations and studies may therefore design for ongoing verification and uncertainty handling, for example by foregrounding evidence coverage for high-level claims, flagging missing or conflicting data, and reinforcing that fluent summaries (even traceable ones) should not be equated with clinical correctness.}

\textbf{Support appropriation by context, experience, and evolving clinical needs. }
Therapists in our study described distinct appropriation patterns based on their clinical context, professional experience level, therapeutic modality, and therapy phases. Recall that expert therapists generally prioritized efficiency, preferring concise summaries for rapid triage, whereas novices relied on detailed summaries and verification features to build confidence and ensure thoroughness. Early therapy stages typically require richer context and detailed reasoning to support comprehensive case formulation, while later stages benefit from succinct, actionable insights to efficiently monitor progress and adapt treatment plans. Additionally, therapists expressed modality-specific information needs, explicitly mentioning assessments like Internal Family Systems (IFS), PTSD Checklist (PCL), and Obsessive-Compulsive Inventory-Revised (OCI-R), underscoring the diversity of clinical approaches.

These findings resonate with longstanding HCI literature emphasizing the importance of flexible, customizable systems that adapt fluidly to diverse practices, experience levels, and situational demands~\cite{dourish2003appropriation, Yuan2025}. Future designs could operationalize such flexibility through adaptive onboarding processes that continually respond to therapists' evolving preferences and therapeutic strategies. For example, therapist-specific profiles and modality-specific assessment packs would enable rapid reconfiguration of data sources, widgets, and summary depth. Lightweight, in-flow customization could allow therapists to effortlessly adjust their configurations as therapeutic needs evolve, shifting seamlessly between detailed views for early-stage assessments and streamlined views for routine monitoring.

Moreover, providing adjustable summary detail sliders (e.g., concise highlights versus detailed views with supporting sources), clinician/client display modes, and experience-level presets (validation-oriented for novices versus efficiency-focused for experts) could further support therapists’ varied workflows and confidence levels. Such adaptive approaches not only enhance the relevance and usability of therapeutic insights but also minimize disruption, promoting sustainable, long-term integration into clinical practice. Longitudinal studies should further investigate how these appropriation behaviors evolve over time, examining their impacts on therapeutic outcomes and clinical workflows.

\subsection{Ethical Considerations of Therapist-AI-interaction}

Our findings surfaced opportunities and risks specific to AI in therapist-facing tools. Below, we outline a few ethical themes and their implications for the design and deployment of future therapist-centered AI tools.

\textbf{Human oversight and accountability.}
Participants consistently framed \tool as an aid, emphasizing that therapeutic judgment must remain solely with the therapist. They appreciated that \tool’s AI-generated summaries supported their interpretation by highlighting relevant patterns and contextual cues, without crossing into definitive judgments or unsolicited therapeutic next steps, which they viewed as intrusive and incompatible with the ethos of psychotherapy. Therapists emphasized that AI support must not override their interpretive authority. Therapists also underscored the need to explicitly mark AI-generated insights as suggestions rather than decisions, thus reinforcing their own clinical responsibility. Our finding aligns with emerging research advocating for human oversight and accountable use of AI in high-stakes scenarios, such as mental health~\cite{amershi2019guidelines, buccinca2021trust}. Future work should further explore best approaches for labeling AI outputs, ensuring direct links back to source data, and prompting therapists to verify key insights before acting upon them for therapists.

\textbf{Avoid unintended harms in therapeutic relationships.}
Therapists valued \tool’s approach of providing concise, AI-generated summaries while preserving access to full client reflections. However, they also raised concerns about potentially negative emotional impacts of presenting AI-derived analytics directly to clients, such as graphs highlighting incomplete homework that might inadvertently induce shame or anxiety. Participants thus emphasized careful control over what insights are displayed and how they are communicated during sessions. This echoes concerns in prior HCI work about designing personal informatics and visualization tools sensitively to avoid unintended emotional harms~\cite{do2023s, franklin2015unintended, vernon1979unintended}. To address this, designers should consider creating separate clinician and client-facing views, providing clear controls for therapists to moderate what data is shared, and enabling customizations aligned explicitly with therapeutic goals (e.g., mood versus sleep tracking or journaling versus exercise completion). Therapists also emphasized that transparency with clients is essential. Clients should be informed when AI is involved in their care and how it is being used. Such openness helps prevent erosion of trust that could arise if clients later discover their therapist had relied on algorithmic summaries without disclosure. Professional guidelines echo this expectation, recommending that clinicians obtain informed consent for AI-assisted interventions and clearly explain what data is collected, how it is processed, and for what purposes~\cite{schwarze_boyd_2025}.

\hl{Moreover, participants pointed out that client acceptability cannot be assumed. Therapists appreciated the potential to offer clients visual explanations, while also recognizing the possibility that clients might view algorithmic involvement as intrusive or misaligned with their expectations for relational care. These reflections reinforce that client-facing safety has to ensure that AI participation aligns with clients' values, preferences, and boundaries. For example, in real-world deployment, clients would reasonably expect their therapeutic data to be processed through closed, institutionally governed models rather than open, general-purpose services. Our study focused on therapists' experiences and did not directly capture clients’ perspectives on AI-augmented homework review. Future research should examine how different client groups perceive the appropriateness of AI involvement in mental health care, what forms of explanations and consent they expect, and how assurances around model governance and closed-system use influence their willingness to engage with such tools.}

\textbf{Privacy, consent, and data minimization.}
Given that therapeutic homework and client journals often contain highly sensitive information, therapists stressed the importance of careful handling of client data in accordance with privacy and security standards (e.g., HIPAA compliance). Participants favored a clear, opt-in approach to data collection and explicit announcement of AI roles, purposes, data use, and retention practices. Our findings resonate with established ethical design principles in HCI concerning informed consent, transparency, and data minimization~\cite{Jakobi2023, Schwind2025, Zhang2024}. Specifically, designers should prioritize transparent consent procedures clearly outlining how client-generated data (e.g., journals, mood logs, biometric inputs) are analyzed by AI, and for what therapeutic purposes. Additionally, tools must provide intuitive controls allowing therapists and clients to restrict or delete sensitive data, ensuring trust and autonomy remain central in the therapist–client relationship.

In summary, dealing with subjective mental health data demands flexible, context-aware AI interfaces that present insights with humility, inviting the clinician’s interpretation, and protecting the client's feelings and dignity.

\section{Limitations}

Our study’s findings should be interpreted in light of several limitations. First, our study was based on a relatively small sample of therapists, which may limit the generalizability of our results to broader clinical populations or different therapy modalities. For example, our sample size may not fully capture the wide variance in individual therapeutic workflows, potentially obscuring specific friction points or adoption barriers that could emerge in more diverse clinical settings. Our participants may also represent a more technology-oriented segment of the population, which could mean our findings underestimate the challenges faced by clinicians with lower digital literacy. Relatedly, the current design of \tool and underlying AI models were tailored to specific homework tracking scenarios and may require adaptation for other therapeutic contexts or populations. Future work should address these limitations through more diverse sampling, longitudinal field deployment, and inclusion of both therapist and client perspectives. \hl{Additionally, while the color scheme used in \mbox{\tool} was intended only for categorical distinction, future iterations should explore more visual and color encodings in visualization design.} Moreover, participants interacted with \tool within one session; real-world adoption and long-term use may reveal additional usability challenges or changes in workflow integration that were not captured here. Additionally, we focused on therapists' perspectives rather than involving clients at this stage, to explore future-oriented AI-powered tools in therapeutic homework management for the safety of AI system deployment. More work is needed to investigate clients' usage patterns, which could provide additional insights. 

\section{Conclusion}
In collaboration with therapists, we designed and developed \tool, a therapist-facing web application to support the review and interpretation of client homework. Through a pilot study with 14 therapists, our findings demonstrate the potential of \tool to support therapists in efficiently tracking and synthesizing client homework. By foregrounding therapist needs and emphasizing transparency, our design helps clinicians conserve cognitive resources, connect patterns in client progress, and maintain professional oversight. While our findings highlight promising avenues for augmenting clinical workflow, we also emphasize the importance of trust calibration and ethical integration of AI in sensitive domains. Looking ahead, we encourage further research to examine long-term adoption, broader user perspectives, and the adaptation of such tools to diverse therapeutic practices. We see therapist-centered, ethically aligned AI tools as a meaningful step toward enhancing human expertise in mental health care.

\begin{acks}
We thank our participants for their time and valuable feedback. We also thank our anonymous reviewers for their reviews. This work is supported by the National Science Foundation under award no. NSF-2517316.
\end{acks}

\bibliographystyle{ACM-Reference-Format}
\bibliography{ref}

@misc{blueprint, 
    title={AI Documentation \& Insights}, 
    url={http://blueprint.ai}, 
    journal={Blueprint.ai}, 
    author={Blueprint for Therapists}, 
    year={2025} 
}

@misc{upheal, 
    title={AI Progress Notes}, 
    url={https://www.upheal.io/?r=0&_gl=1}, 
    journal={Upheal.io}, 
    author={Upheal}, 
    year={2025} 
}

@article{zhang2018idmvis,
    author = {Zhang, Yixuan and Chanana, Kartik and Dunne, Cody},
    title = {IDMVis: Temporal Event Sequence Visualization for Type 1 Diabetes Treatment Decision Support},
    year = {2019},
    issue_date = {Jan. 2019},
    publisher = {IEEE Educational Activities Department},
    address = {USA},
    volume = {25},
    number = {1},
    issn = {1077-2626},
    url = {https://doi.org/10.1109/TVCG.2018.2865076},
    doi = {10.1109/TVCG.2018.2865076},
    journal = {IEEE Transactions on Visualization and Computer Graphics},
    month = jan,
    pages = {512–522},
    numpages = {11}
}

@inproceedings{amershi2019guidelines,
    author = {Amershi, Saleema and Weld, Dan and Vorvoreanu, Mihaela and Fourney, Adam and Nushi, Besmira and Collisson, Penny and Suh, Jina and Iqbal, Shamsi and Bennett, Paul N. and Inkpen, Kori and Teevan, Jaime and Kikin-Gil, Ruth and Horvitz, Eric},
    title = {Guidelines for Human-AI Interaction},
    year = {2019},
    isbn = {9781450359702},
    publisher = {Association for Computing Machinery},
    address = {New York, NY, USA},
    url = {https://doi.org/10.1145/3290605.3300233},
    doi = {10.1145/3290605.3300233},
    booktitle = {Proceedings of the 2019 CHI Conference on Human Factors in Computing Systems},
    pages = {1–13},
    numpages = {13},
    location = {Glasgow, Scotland Uk},
    series = {CHI '19}
}

@inproceedings{liao2020questioning,
    author = {Liao, Q. Vera and Gruen, Daniel and Miller, Sarah},
    title = {Questioning the AI: Informing Design Practices for Explainable AI User Experiences},
    year = {2020},
    isbn = {9781450367080},
    publisher = {Association for Computing Machinery},
    address = {New York, NY, USA},
    url = {https://doi.org/10.1145/3313831.3376590},
    doi = {10.1145/3313831.3376590},
    booktitle = {Proceedings of the 2020 CHI Conference on Human Factors in Computing Systems},
    pages = {1–15},
    numpages = {15},
    location = {Honolulu, HI, USA},
    series = {CHI '20}
}

@article{buccinca2021trust,
    author = {Bu\c{c}inca, Zana and Malaya, Maja Barbara and Gajos, Krzysztof Z.},
    title = {To Trust or to Think: Cognitive Forcing Functions Can Reduce Overreliance on AI in AI-assisted Decision-making},
    year = {2021},
    issue_date = {April 2021},
    publisher = {Association for Computing Machinery},
    address = {New York, NY, USA},
    volume = {5},
    number = {CSCW1},
    url = {https://doi.org/10.1145/3449287},
    doi = {10.1145/3449287},
    journal = {Proc. ACM Hum.-Comput. Interact.},
    month = apr,
    articleno = {188},
    numpages = {21}
}

@inproceedings{bansal2021,
    author = {Bansal, Gagan and Wu, Tongshuang and Zhou, Joyce and Fok, Raymond and Nushi, Besmira and Kamar, Ece and Ribeiro, Marco Tulio and Weld, Daniel},
    title = {Does the Whole Exceed its Parts? The Effect of AI Explanations on Complementary Team Performance},
    year = {2021},
    isbn = {9781450380966},
    publisher = {Association for Computing Machinery},
    address = {New York, NY, USA},
    url = {https://doi.org/10.1145/3411764.3445717},
    doi = {10.1145/3411764.3445717},
    booktitle = {Proceedings of the 2021 CHI Conference on Human Factors in Computing Systems},
    articleno = {81},
    numpages = {16},
    location = {Yokohama, Japan},
    series = {CHI '21}
}

@inproceedings{cai2019effects,
    author = {Cai, Carrie J. and Reif, Emily and Hegde, Narayan and Hipp, Jason and Kim, Been and Smilkov, Daniel and Wattenberg, Martin and Viegas, Fernanda and Corrado, Greg S. and Stumpe, Martin C. and Terry, Michael},
    title = {Human-Centered Tools for Coping with Imperfect Algorithms During Medical Decision-Making},
    year = {2019},
    isbn = {9781450359702},
    publisher = {Association for Computing Machinery},
    address = {New York, NY, USA},
    url = {https://doi.org/10.1145/3290605.3300234},
    doi = {10.1145/3290605.3300234},
    booktitle = {Proceedings of the 2019 CHI Conference on Human Factors in Computing Systems},
    pages = {1–14},
    numpages = {14},
    location = {Glasgow, Scotland Uk},
    series = {CHI '19}
}

@inproceedings{shneiderman2003eyes,
    author = {Shneiderman, Ben},
    title = {The Eyes Have It: A Task by Data Type Taxonomy for Information Visualizations},
    year = {1996},
    isbn = {081867508X},
    publisher = {IEEE Computer Society},
    address = {USA},
    booktitle = {Proceedings of the 1996 IEEE Symposium on Visual Languages},
    pages = {336},
    series = {VL '96}
}

@inproceedings{kulesza2015principles,
    author = {Kulesza, Todd and Burnett, Margaret and Wong, Weng-Keen and Stumpf, Simone},
    title = {Principles of Explanatory Debugging to Personalize Interactive Machine Learning},
    year = {2015},
    isbn = {9781450333061},
    publisher = {Association for Computing Machinery},
    address = {New York, NY, USA},
    url = {https://doi.org/10.1145/2678025.2701399},
    doi = {10.1145/2678025.2701399},
    booktitle = {Proceedings of the 20th International Conference on Intelligent User Interfaces},
    pages = {126–137},
    numpages = {12},
    location = {Atlanta, Georgia, USA},
    series = {IUI '15}
}

@article{dourish2003appropriation,
    author = {Dourish, Paul},
    title = {The Appropriation of Interactive Technologies: Some Lessons from Placeless Documents},
    year = {2003},
    issue_date = {2003},
    publisher = {Kluwer Academic Publishers},
    address = {USA},
    volume = {12},
    number = {4},
    issn = {0925-9724},
    url = {https://doi.org/10.1023/A:1026149119426},
    doi = {10.1023/A:1026149119426},
    journal = {Comput. Supported Coop. Work},
    month = sep,
    pages = {465–490},
    numpages = {26}
}

@inbook{franklin2015unintended,
  title = {The Unintended Consequences of the Technology in Clinical Settings},
  ISBN = {9783319172729},
  ISSN = {2197-3741},
  url = {http://dx.doi.org/10.1007/978-3-319-17272-9_11},
  DOI = {10.1007/978-3-319-17272-9_11},
  booktitle = {Cognitive Informatics for Biomedicine},
  publisher = {Springer International Publishing},
  author = {Franklin,  Amy},
  year = {2015},
  pages = {241–258}
}

@article{vernon1979unintended, title={Unintended Consequences}, volume={7}, url={https://www.jstor.org/stable/190824}, number={1}, journal={Political Theory}, author={Vernon, Richard}, year={1979}, pages={57–73} }

@inproceedings{do2023s,
    author = {Do, Kimberly and Pang, Rock Yuren and Jiang, Jiachen and Reinecke, Katharina},
    title = {“That’s important, but...”: How Computer Science Researchers Anticipate Unintended Consequences of Their Research Innovations},
    year = {2023},
    isbn = {9781450394215},
    publisher = {Association for Computing Machinery},
    address = {New York, NY, USA},
    url = {https://doi.org/10.1145/3544548.3581347},
    doi = {10.1145/3544548.3581347},
    booktitle = {Proceedings of the 2023 CHI Conference on Human Factors in Computing Systems},
    articleno = {602},
    numpages = {16},
    location = {Hamburg, Germany},
    series = {CHI '23}
}

@inproceedings{Fallman2003,
    author = {Fallman, Daniel},
    title = {Design-oriented human-computer interaction},
    year = {2003},
    isbn = {1581136307},
    publisher = {Association for Computing Machinery},
    address = {New York, NY, USA},
    url = {https://doi.org/10.1145/642611.642652},
    doi = {10.1145/642611.642652},
    booktitle = {Proceedings of the SIGCHI Conference on Human Factors in Computing Systems},
    pages = {225–232},
    numpages = {8},
    location = {Ft. Lauderdale, Florida, USA},
    series = {CHI '03}
}

@article{Prasko2022,
  title = {Homework in Cognitive Behavioral Supervision: Theoretical Background and Clinical Application},
  volume = {Volume 15},
  ISSN = {1179-1578},
  url = {http://dx.doi.org/10.2147/PRBM.S382246},
  DOI = {10.2147/prbm.s382246},
  journal = {Psychology Research and Behavior Management},
  publisher = {Informa UK Limited},
  author = {Prasko,  Jan and Krone,  Ilona and Burkauskas,  Julius and Vanek,  Jakub and Abeltina,  Marija and Juskiene,  Alicja and Sollar,  Tomas and Bite,  Ieva and Slepecky,  Milos and Ociskova,  Marie},
  year = {2022},
  month = dec,
  pages = {3809–3824}
}

@article{Freeman2007,
  title = {The Use of Homework in Cognitive Behavior Therapy: Working with Complex Anxiety and Insomnia},
  volume = {14},
  ISSN = {1077-7229},
  url = {http://dx.doi.org/10.1016/j.cbpra.2006.10.005},
  DOI = {10.1016/j.cbpra.2006.10.005},
  number = {3},
  journal = {Cognitive and Behavioral Practice},
  publisher = {Elsevier BV},
  author = {Freeman,  Arthur},
  year = {2007},
  month = aug,
  pages = {261–267}
}

@article{Kazantzis2010,
  title = {Meta-analysis of homework effects in cognitive and behavioral therapy: A replication and extension.},
  volume = {17},
  ISSN = {0969-5893},
  url = {http://dx.doi.org/10.1111/j.1468-2850.2010.01204.x},
  DOI = {10.1111/j.1468-2850.2010.01204.x},
  number = {2},
  journal = {Clinical Psychology: Science and Practice},
  publisher = {American Psychological Association (APA)},
  author = {Kazantzis,  Nikolaos and Whittington,  Craig and Dattilio,  Frank},
  year = {2010},
  month = jun,
  pages = {144–156}
}

@article{Mausbach2010,
  title = {The Relationship Between Homework Compliance and Therapy Outcomes: An Updated Meta-Analysis},
  volume = {34},
  ISSN = {1573-2819},
  url = {http://dx.doi.org/10.1007/s10608-010-9297-z},
  DOI = {10.1007/s10608-010-9297-z},
  number = {5},
  journal = {Cognitive Therapy and Research},
  publisher = {Springer Science and Business Media LLC},
  author = {Mausbach,  Brent T. and Moore,  Raeanne and Roesch,  Scott and Cardenas,  Veronica and Patterson,  Thomas L.},
  year = {2010},
  month = feb,
  pages = {429–438}
}

@article{Horvath2011,
  title = {Alliance in individual psychotherapy.},
  volume = {48},
  ISSN = {0033-3204},
  url = {http://dx.doi.org/10.1037/a0022186},
  DOI = {10.1037/a0022186},
  number = {1},
  journal = {Psychotherapy},
  publisher = {American Psychological Association (APA)},
  author = {Horvath,  Adam O. and Del Re,  A. C. and Fl\"{u}ckiger,  Christoph and Symonds,  Dianne},
  year = {2011},
  pages = {9–16}
}

@article{Scheel2004,
  title = {The Process of Recommending Homework in Psychotherapy: A Review of Therapist Delivery Methods,  Client Acceptability,  and Factors That Affect Compliance.},
  volume = {41},
  ISSN = {0033-3204},
  url = {http://dx.doi.org/10.1037/0033-3204.41.1.38},
  DOI = {10.1037/0033-3204.41.1.38},
  number = {1},
  journal = {Psychotherapy: Theory,  Research,  Practice,  Training},
  publisher = {American Psychological Association (APA)},
  author = {Scheel,  Michael J. and Hanson,  William E. and Razzhavaikina,  Tanya I.},
  year = {2004},
  pages = {38–55}
}

@article{Bunnell2024,
  title = {Expanding a Health Technology Solution to Address Therapist Challenges in Implementing Homework With Adult Clients: Mixed Methods Study},
  volume = {11},
  ISSN = {2292-9495},
  url = {http://dx.doi.org/10.2196/56567},
  DOI = {10.2196/56567},
  journal = {JMIR Human Factors},
  publisher = {JMIR Publications Inc.},
  author = {Bunnell,  Brian E and Schuler,  Kaitlyn R and Ivanova,  Julia and Flynn,  Lea and Barrera,  Janelle F and Niazi,  Jasmine and Turner,  Dylan and Welch,  Brandon M},
  year = {2024},
  month = dec,
  pages = {e56567}
}

@article{Dattilio2011,
  title = {A Survey of Homework Use,  Experience of Barriers to Homework,  and Attitudes About the Barriers to Homework Among Couples and Family Therapists},
  volume = {37},
  ISSN = {1752-0606},
  url = {http://dx.doi.org/10.1111/j.1752-0606.2011.00223.x},
  DOI = {10.1111/j.1752-0606.2011.00223.x},
  number = {2},
  journal = {Journal of Marital and Family Therapy},
  publisher = {Wiley},
  author = {Dattilio,  Frank M. and Kazantzis,  Nikolaos and Shinkfield,  Gregg and Carr,  Amanda G.},
  year = {2011},
  month = mar,
  pages = {121–136}
}

@book{cully_dawson_hamer_tharp_2020, 
    address={Houston, TX}, 
    title={A Provider’s Guide to Brief Cognitive Behavioral Therapy}, 
    url={https://www.mirecc.va.gov/visn16/docs/therapists_guide_to_brief_cbtmanual.pdf?utm_source=chatgpt.com}, 
    publisher={Department of Veterans Affairs South Central MIRECC}, 
    author={Cully, Jeffrey A. and Dawson, Darius B. and Hamer, Joshua and Tharp, Andra Teten}, year={2020} 
}

@inproceedings{Perttu2023,
    author = {H\"{a}m\"{a}l\"{a}inen, Perttu and Tavast, Mikke and Kunnari, Anton},
    title = {Evaluating Large Language Models in Generating Synthetic HCI Research Data: a Case Study},
    year = {2023},
    isbn = {9781450394215},
    publisher = {Association for Computing Machinery},
    address = {New York, NY, USA},
    url = {https://doi.org/10.1145/3544548.3580688},
    doi = {10.1145/3544548.3580688},
    booktitle = {Proceedings of the 2023 CHI Conference on Human Factors in Computing Systems},
    articleno = {433},
    numpages = {19},
    location = {Hamburg, Germany},
    series = {CHI '23}
}

@inproceedings{Kambhamettu2024,
    author = {Kambhamettu, Hita and Metaxa, Dana\"{e} and Johnson, Kevin and Head, Andrew},
    title = {Explainable Notes: Examining How to Unlock Meaning in Medical Notes with Interactivity and Artificial Intelligence},
    year = {2024},
    isbn = {9798400703300},
    publisher = {Association for Computing Machinery},
    address = {New York, NY, USA},
    url = {https://doi.org/10.1145/3613904.3642573},
    doi = {10.1145/3613904.3642573},
    booktitle = {Proceedings of the 2024 CHI Conference on Human Factors in Computing Systems},
    articleno = {449},
    numpages = {19},
    location = {Honolulu, HI, USA},
    series = {CHI '24}
}

@article{Biswas2024,
  title = {Intelligent Clinical Documentation: Harnessing Generative AI for Patient-Centric Clinical Note Generation},
  ISSN = {2456-2165},
  url = {http://dx.doi.org/10.38124/ijisrt/IJISRT24MAY1483},
  DOI = {10.38124/ijisrt/ijisrt24may1483},
  journal = {International Journal of Innovative Science and Research Technology (IJISRT)},
  publisher = {International Journal of Innovative Science and Research Technology},
  author = {Biswas,  Anjanava and Talukdar,  Wrick},
  year = {2024},
  month = may,
  pages = {994–1008}
}

@inproceedings{Kim2024,
  series = {CHI ’24},
  title = {MindfulDiary: Harnessing Large Language Model to Support Psychiatric Patients’ Journaling},
  url = {http://dx.doi.org/10.1145/3613904.3642937},
  DOI = {10.1145/3613904.3642937},
  booktitle = {Proceedings of the CHI Conference on Human Factors in Computing Systems},
  publisher = {ACM},
  author = {Kim,  Taewan and Bae,  Seolyeong and Kim,  Hyun Ah and Lee,  Su-Woo and Hong,  Hwajung and Yang,  Chanmo and Kim,  Young-Ho},
  year = {2024},
  month = may,
  pages = {1–20},
  collection = {CHI ’24}
}

@inproceedings{Nepal2024,
    author = {Nepal, Subigya and Pillai, Arvind and Campbell, William and Massachi, Talie and Choi, Eunsol Soul and Xu, Xuhai and Kuc, Joanna and Huckins, Jeremy F and Holden, Jason and Depp, Colin and Jacobson, Nicholas and Czerwinski, Mary P and Granholm, Eric and Campbell, Andrew},
    title = {Contextual AI Journaling: Integrating LLM and Time Series Behavioral Sensing Technology to Promote Self-Reflection and Well-being using the MindScape App},
    year = {2024},
    isbn = {9798400703317},
    publisher = {Association for Computing Machinery},
    address = {New York, NY, USA},
    url = {https://doi.org/10.1145/3613905.3650767},
    doi = {10.1145/3613905.3650767},
    booktitle = {Extended Abstracts of the CHI Conference on Human Factors in Computing Systems},
    articleno = {86},
    numpages = {8},
    location = {Honolulu, HI, USA},
    series = {CHI EA '24}
}

@article{Wang2025,
  title = {Evaluating Generative AI in Mental Health: Systematic Review of Capabilities and Limitations},
  volume = {12},
  ISSN = {2368-7959},
  url = {http://dx.doi.org/10.2196/70014},
  DOI = {10.2196/70014},
  journal = {JMIR Mental Health},
  publisher = {JMIR Publications Inc.},
  author = {Wang,  Liying and Bhanushali,  Tanmay and Huang,  Zhuoran and Yang,  Jingyi and Badami,  Sukriti and Hightow-Weidman,  Lisa},
  year = {2025},
  month = may,
  pages = {e70014–e70014}
}

@article{Siddals2024,
  title = {“It happened to be the perfect thing”: experiences of generative AI chatbots for mental health},
  volume = {3},
  ISSN = {2731-4251},
  url = {http://dx.doi.org/10.1038/s44184-024-00097-4},
  DOI = {10.1038/s44184-024-00097-4},
  number = {1},
  journal = {npj Mental Health Research},
  publisher = {Springer Science and Business Media LLC},
  author = {Siddals,  Steven and Torous,  John and Coxon,  Astrid},
  year = {2024},
  month = oct 
}

@article{Stade2024,
  title = {Large language models could change the future of behavioral healthcare: a proposal for responsible development and evaluation},
  volume = {3},
  ISSN = {2731-4251},
  url = {http://dx.doi.org/10.1038/s44184-024-00056-z},
  DOI = {10.1038/s44184-024-00056-z},
  number = {1},
  journal = {npj Mental Health Research},
  publisher = {Springer Science and Business Media LLC},
  author = {Stade,  Elizabeth C. and Stirman,  Shannon Wiltsey and Ungar,  Lyle H. and Boland,  Cody L. and Schwartz,  H. Andrew and Yaden,  David B. and Sedoc,  João and DeRubeis,  Robert J. and Willer,  Robb and Eichstaedt,  Johannes C.},
  year = {2024},
  month = apr 
}

@misc{qualtrics_2025,
    title={The Leading Research \& Experience Software | Qualtrics}, 
    url={https://www.qualtrics.com/}, 
    author={Qualtrics}, 
    year={2025} 
}

@article{thomas2006general,
  title={A general inductive approach for analyzing qualitative evaluation data},
  author={Thomas, David R},
  journal={American journal of evaluation},
  volume={27},
  number={2},
  pages={237--246},
  year={2006},
  publisher={Sage Publications Sage CA: Thousand Oaks, CA}
}

@misc{NextJS,
  author = {Vercel},
  title = {Next.js: The React Framework for Production},
  url = {https://nextjs.org},
  year = {2024}
}

@misc{microsoft_azure_openai,
  author = {Microsoft},
  title = {Azure OpenAI Service},
  url = {https://learn.microsoft.com/en-us/azure/ai-services/openai/},
  year = {2024}
}

@misc{langchain,
    author = {LangChain}, 
    title={LangChain: Building applications with LLMs through composability},
    url={https://www.langchain.com/},
    year = {2024}
}

@article{Schnall2018,
  author = {Schnall, Rebecca and Cho, Hwayoung and Liu, Jiantao},
  title = {Health Information Technology Usability Evaluation Scale (Health-ITUES) for Usability Assessment of Mobile Health Technology: Validation Study},
  journal = {JMIR Mhealth Uhealth},
  volume = {6},
  number = {1},
  pages = {e4},
  year = {2018},
  pmid = {29305343},
  doi = {10.2196/mhealth.8048}
}

@inproceedings{WangY2025,
    author = {Wang, Yimeng and Wang, Yinzhou and Crace, Kelly and Zhang, Yixuan},
    title = {Understanding Attitudes and Trust of Generative AI Chatbots for Social Anxiety Support},
    year = {2025},
    isbn = {9798400713941},
    publisher = {Association for Computing Machinery},
    address = {New York, NY, USA},
    url = {https://doi.org/10.1145/3706598.3714286},
    doi = {10.1145/3706598.3714286},
    booktitle = {Proceedings of the 2025 CHI Conference on Human Factors in Computing Systems},
    articleno = {1123},
    numpages = {21},
    series = {CHI '25}
}

@article{Bordin1979,
  title = {The generalizability of the psychoanalytic concept of the working alliance.},
  volume = {16},
  ISSN = {0033-3204},
  url = {http://dx.doi.org/10.1037/h0085885},
  DOI = {10.1037/h0085885},
  number = {3},
  journal = {Psychotherapy: Theory,  Research \& Practice},
  publisher = {American Psychological Association (APA)},
  author = {Bordin,  Edward S.},
  year = {1979},
  pages = {252–260}
}

@book{schwarze_boyd_2025, 
    title={Best Practices for the Use of Artificial Intelligence by Mental Health Therapists}, url={https://ai.utah.gov/wp-content/uploads/Best-Practices-Mental-Health-Therapists.pdf?utm_source=chatgpt.com},
    publisher={Utah Office of Artificial Intelligence Policy (OAIP) and the Utah Division of Professional Licensing (DOPL)},
    author={Schwarze, Alice C. and Boyd, Zachary M. },
    year={2025},
    month={Apr}
}

@inbook{Jakobi2023,
  title = {What HCI Can Do for (Data Protection) Law—Beyond Design},
  ISBN = {9783031286438},
  url = {http://dx.doi.org/10.1007/978-3-031-28643-8_6},
  DOI = {10.1007/978-3-031-28643-8_6},
  booktitle = {Human Factors in Privacy Research},
  publisher = {Springer International Publishing},
  author = {Jakobi,  Timo and Grafenstein,  Maximilian von},
  year = {2023},
  pages = {115–136}
}

@article{Schwind2025,
    author = {Schwind, Valentin and Zelalem Tadesse, Netsanet and Silva da Cunha, Estefania and Hamidi, Yeganeh and Sanjar Sultani, Soltan and Sehrt, Jessica},
    title = {A Scoping Review of Informed Consent Practices in Human–Computer Interaction Research},
    year = {2025},
    issue_date = {August 2025},
    publisher = {Association for Computing Machinery},
    address = {New York, NY, USA},
    volume = {32},
    number = {4},
    issn = {1073-0516},
    url = {https://doi.org/10.1145/3721284},
    doi = {10.1145/3721284},
    journal = {ACM Trans. Comput.-Hum. Interact.},
    month = aug,
    articleno = {35},
    numpages = {60}
}

@misc{Petrik2015,
  title = {Homework in Psychotherapy},
  ISBN = {9781118625392},
  url = {http://dx.doi.org/10.1002/9781118625392.wbecp081},
  DOI = {10.1002/9781118625392.wbecp081},
  journal = {The Encyclopedia of Clinical Psychology},
  publisher = {Wiley},
  author = {Petrik,  Alexandra M. and Kazantzis,  Nikolaos},
  year = {2015},
  month = jan,
  pages = {1–6}
}

@book{linehan2015dbt,
  author    = {Linehan, Marsha M.},
  title     = {DBT{\textregistered} Skills Training Handouts and Worksheets},
  edition   = {2nd},
  year      = {2015},
  publisher = {The Guilford Press},
  address   = {New York, NY},
}

@article{Jerome2023,
  title = {Solution-focused approaches in adult mental health research: A conceptual literature review and narrative synthesis},
  volume = {14},
  ISSN = {1664-0640},
  url = {http://dx.doi.org/10.3389/fpsyt.2023.1068006},
  DOI = {10.3389/fpsyt.2023.1068006},
  journal = {Frontiers in Psychiatry},
  publisher = {Frontiers Media SA},
  author = {Jerome,  Lauren and McNamee,  Philip and Abdel-Halim,  Nadia and Elliot,  Kathryn and Woods,  Jonathan},
  year = {2023},
  month = mar 
}

@book{miller2002motivational,
  author    = {Miller, William R. and Rollnick, Stephen},
  title     = {Motivational Interviewing: Preparing People for Change},
  edition   = {2nd},
  year      = {2002},
  publisher = {The Guilford Press},
  address   = {New York, NY},
}

@article{hayes2006acceptance,
  author    = {Hayes, Stephen C. and Luoma, Jason B. and Bond, Frank W. and Masuda, Akihiko and Lillis, Jason},
  title     = {Acceptance and Commitment Therapy: Model, Processes and Outcomes},
  journal   = {Psychology Faculty Publications},
  year      = {2006},
  number    = {101},
  url       = {https://scholarworks.gsu.edu/psych_facpub/101},
}

@article{Burns2000,
  title = {Does psychotherapy homework lead to improvements in depression in cognitive–behavioral therapy or does improvement lead to increased homework compliance?},
  volume = {68},
  ISSN = {0022-006X},
  url = {http://dx.doi.org/10.1037/0022-006X.68.1.46},
  DOI = {10.1037/0022-006x.68.1.46},
  number = {1},
  journal = {Journal of Consulting and Clinical Psychology},
  publisher = {American Psychological Association (APA)},
  author = {Burns,  David D. and Spangler,  Diane L.},
  year = {2000},
  pages = {46–56}
}

@article{Conklin2015,
  title = {A session-to-session examination of homework engagement in cognitive therapy for depression: Do patients experience immediate benefits?},
  volume = {72},
  ISSN = {0005-7967},
  url = {http://dx.doi.org/10.1016/j.brat.2015.06.011},
  DOI = {10.1016/j.brat.2015.06.011},
  journal = {Behaviour Research and Therapy},
  publisher = {Elsevier BV},
  author = {Conklin,  Laren R. and Strunk,  Daniel R.},
  year = {2015},
  month = sep,
  pages = {56–62}
}

@article{CamminNowak2013,
  title = {Specificity of Homework Compliance Effects on Treatment Outcome in CBT: Evidence from a Controlled Trial on Panic Disorder and Agoraphobia: Homework Effects in CBT for Panic Disorder},
  volume = {69},
  ISSN = {0021-9762},
  url = {http://dx.doi.org/10.1002/jclp.21975},
  DOI = {10.1002/jclp.21975},
  number = {6},
  journal = {Journal of Clinical Psychology},
  publisher = {Wiley},
  author = {Cammin-Nowak,  Sandra and Helbig-Lang,  Sylvia and Lang,  Thomas and Gloster,  Andrew T. and Fehm,  Lydia and Gerlach,  Alexander L. and Str\"{o}hle,  Andreas and Deckert,  J\"{u}rgen and Kircher,  Tilo and Hamm,  Alfons O. and Alpers,  Georg W. and Arolt,  Volker and Wittchen,  H.-U.},
  year = {2013},
  month = mar,
  pages = {616–629}
}

@article{Cooper2017,
  title = {Homework “Dose, ” Type,  and Helpfulness as Predictors of Clinical Outcomes in Prolonged Exposure for PTSD},
  volume = {48},
  ISSN = {0005-7894},
  url = {http://dx.doi.org/10.1016/j.beth.2016.02.013},
  DOI = {10.1016/j.beth.2016.02.013},
  number = {2},
  journal = {Behavior Therapy},
  publisher = {Elsevier BV},
  author = {Cooper,  Andrew A. and Kline,  Alexander C. and Graham,  Belinda and Bedard-Gilligan,  Michele and Mello,  Patricia G. and Feeny,  Norah C. and Zoellner,  Lori A.},
  year = {2017},
  month = mar,
  pages = {182–194}
}

@article{Carroll2005,
  title = {Practice Makes Progress? Homework Assignments and Outcome in Treatment of Cocaine Dependence.},
  volume = {73},
  ISSN = {0022-006X},
  url = {http://dx.doi.org/10.1037/0022-006X.73.4.749},
  DOI = {10.1037/0022-006x.73.4.749},
  number = {4},
  journal = {Journal of Consulting and Clinical Psychology},
  publisher = {American Psychological Association (APA)},
  author = {Carroll,  Kathleen M. and Nich,  Charla and Ball,  Samuel A.},
  year = {2005},
  month = aug,
  pages = {749–755}
}

@article{korotitsch1999overview,
  author    = {Korotitsch, William J. and Nelson-Gray, Rosemery O.},
  title     = {An Overview of Self-Monitoring Research in Assessment and Treatment},
  journal   = {Psychological Assessment},
  year      = {1999},
  volume    = {11},
  number    = {4},
  pages     = {415--425},
  doi       = {10.1037/1040-3590.11.4.415},
  url       = {https://doi.org/10.1037/1040-3590.11.4.415},
}

@book{linehan2015manual,
  author    = {Linehan, Marsha M.},
  title     = {DBT Skills Training Manual},
  edition   = {2nd},
  year      = {2015},
  publisher = {The Guilford Press},
  address   = {New York, NY},
}

@book{bennettlevy2004oxford,
  editor    = {Bennett-Levy, James and Butler, Gillian and Fennell, Melanie and Hackmann, Ann and Mueller, Martina and Westbrook, David},
  title     = {Oxford Guide to Behavioural Experiments in Cognitive Therapy},
  year      = {2004},
  publisher = {Oxford University Press},
  address   = {Oxford, UK},
}

@book{martell2022behavioral,
  author    = {Martell, Christopher R. and Dimidjian, Sona and Herman-Dunn, Ruth},
  title     = {Behavioral Activation for Depression: A Clinician's Guide},
  edition   = {2nd},
  year      = {2022},
  publisher = {The Guilford Press},
  address   = {New York, NY},
  isbn      = {9781462548385},
}

@book{hayes2011acceptance,
  author    = {Hayes, Steven C. and Strosahl, Kirk D. and Wilson, Kelly G.},
  title     = {Acceptance and Commitment Therapy, Second Edition: The Process and Practice of Mindful Change},
  edition   = {2nd},
  year      = {2011},
  publisher = {The Guilford Press},
  address   = {New York, NY},
}

@online{arttherapyresources2023session,
  author    = {Art Therapy},
  title     = {How to Begin and End an Art Therapy Session Effectively},
  year      = {2023},
  url       = {https://arttherapyresources.com.au/art-therapy-session-guide/},
  note      = {Accessed: 2025-09-08},
}

@article{Moberg2019,
  title = {Guided Self-Help Works: Randomized Waitlist Controlled Trial of Pacifica,  a Mobile App Integrating Cognitive Behavioral Therapy and Mindfulness for Stress,  Anxiety,  and Depression},
  volume = {21},
  ISSN = {1438-8871},
  url = {http://dx.doi.org/10.2196/12556},
  DOI = {10.2196/12556},
  number = {6},
  journal = {Journal of Medical Internet Research},
  publisher = {JMIR Publications Inc.},
  author = {Moberg,  Christine and Niles,  Andrea and Beermann,  Dale},
  year = {2019},
  month = jun,
  pages = {e12556}
}

@article{ODaffer2022,
  title = {Efficacy and Conflicts of Interest in Randomized Controlled Trials Evaluating Headspace and Calm Apps: Systematic Review},
  volume = {9},
  ISSN = {2368-7959},
  url = {http://dx.doi.org/10.2196/40924},
  DOI = {10.2196/40924},
  number = {9},
  journal = {JMIR Mental Health},
  publisher = {JMIR Publications Inc.},
  author = {O’Daffer,  Alison and Colt,  Susannah F and Wasil,  Akash R and Lau,  Nancy},
  year = {2022},
  month = sep,
  pages = {e40924}
}

@article{Lewis2019,
  title = {Implementing Measurement-Based Care in Behavioral Health: A Review},
  volume = {76},
  ISSN = {2168-622X},
  url = {http://dx.doi.org/10.1001/jamapsychiatry.2018.3329},
  DOI = {10.1001/jamapsychiatry.2018.3329},
  number = {3},
  journal = {JAMA Psychiatry},
  publisher = {American Medical Association (AMA)},
  author = {Lewis,  Cara C. and Boyd,  Meredith and Puspitasari,  Ajeng and Navarro,  Elena and Howard,  Jacqueline and Kassab,  Hannah and Hoffman,  Mira and Scott,  Kelli and Lyon,  Aaron and Douglas,  Susan and Simon,  Greg and Kroenke,  Kurt},
  year = {2019},
  month = mar,
  pages = {324}
}

@misc{MeyerKalos2024,
  title = {Putting measurement-based care into action: A mixed methods study of the benefits of integrating routine client feedback in coordinated specialty care programs for early psychosis},
  url = {http://dx.doi.org/10.21203/rs.3.rs-3918063/v1},
  DOI = {10.21203/rs.3.rs-3918063/v1},
  publisher = {Springer Science and Business Media LLC},
  author = {Meyer-Kalos,  Piper and Owens,  Grace and Fisher,  Melissa and Wininger,  Lionel and Williams-Wengerd,  Anne and Breen,  Kimberleigh and Abate,  Josephine and Currie,  Ariel and Olinger,  Nathan and Vinogradov,  Sophia},
  year = {2024},
  month = feb 
}

@article{Ridout2025,
  title = {Considerations for Implementation of Measurement-Based Care: Focus on Solo and Small-Group Practitioners},
  volume = {76},
  ISSN = {1557-9700},
  url = {http://dx.doi.org/10.1176/appi.ps.20240372},
  DOI = {10.1176/appi.ps.20240372},
  number = {7},
  journal = {Psychiatric Services},
  publisher = {American Psychiatric Association Publishing},
  author = {Ridout,  Kathryn K. and Vanderlip,  Erik and Carlo,  Andrew D. and Kadriu,  Bashkim and Livesey,  Cecilia and Torous,  John and Alter,  Carol},
  year = {2025},
  month = jul,
  pages = {665–674}
}

@misc{mirah_website,
  author = {Mirah},
  title  = {Powering Behavioral Health Integration},
  year   = {2025},
  note   = {Accessed: 2025-09-08},
  url    = {https://www.mirah.com/},
}

@article{Chang2024,
  title = {AI-Led Mental Health Support (Wysa) for Health Care Workers During COVID-19: Service Evaluation},
  volume = {8},
  ISSN = {2561-326X},
  url = {http://dx.doi.org/10.2196/51858},
  DOI = {10.2196/51858},
  journal = {JMIR Formative Research},
  publisher = {JMIR Publications Inc.},
  author = {Chang,  Christel Lynne and Sinha,  Chaitali and Roy,  Madhavi and Wong,  John Chee Meng},
  year = {2024},
  month = apr,
  pages = {e51858}
}

@article{Mehta2021,
  title = {Acceptability and Effectiveness of Artificial Intelligence Therapy for Anxiety and Depression (Youper): Longitudinal Observational Study},
  volume = {23},
  ISSN = {1438-8871},
  url = {http://dx.doi.org/10.2196/26771},
  DOI = {10.2196/26771},
  number = {6},
  journal = {Journal of Medical Internet Research},
  publisher = {JMIR Publications Inc.},
  author = {Mehta,  Ashish and Niles,  Andrea Nicole and Vargas,  Jose Hamilton and Marafon,  Thiago and Couto,  Diego Dotta and Gross,  James Jonathan},
  year = {2021},
  month = jun,
  pages = {e26771}
}

@misc{headspace,
  author       = {Headspace Inc.},
  title        = {Headspace: Mindful meditation},
  howpublished = {Mobile application},
  year         = {2024},
  url          = {https://www.headspace.com/}
}

@misc{SanvelloApp,
    author = {Sanvello Health, Inc.},
    title = {Sanvello},
    url  =  {https://www.sanvello.com/},
    year = {2024}
}

@article{inkster2018empathy,
  author = {Inkster, Becky and Sarda, Shubhankar and Subramanian, Vinod},
  title = {An Empathy-Driven, Conversational Artificial Intelligence Agent ({Wysa}) for Digital Mental Well-Being: Real-World Data Evaluation Mixed-Methods Study},
  journal = {JMIR mHealth and uHealth},
  volume = {6},
  number = {11},
  pages = {e12106},
  year = {2018},
  doi = {10.2196/12106},
  url = {https://mhealth.jmir.org/2018/11/e12106/}
}

@software{youper,
  author = {Youper Inc.},
  title = {Youper: AI Mental Health},
  url = {https://www.youper.ai/},
  year = {2016},
  note = {Mobile app},
  organization = {Youper Inc.},
  address = {San Francisco, CA}
}

@article{Kernberg2024,
  title = {Using ChatGPT-4 to Create Structured Medical Notes From Audio Recordings of Physician-Patient Encounters: Comparative Study},
  volume = {26},
  ISSN = {1438-8871},
  url = {http://dx.doi.org/10.2196/54419},
  DOI = {10.2196/54419},
  journal = {Journal of Medical Internet Research},
  publisher = {JMIR Publications Inc.},
  author = {Kernberg,  Annessa and Gold,  Jeffrey A and Mohan,  Vishnu},
  year = {2024},
  month = apr,
  pages = {e54419}
}

@article{Lee2024,
  title = {Evaluating the Impact of Artificial Intelligence (AI) on Clinical Documentation Efficiency and Accuracy Across Clinical Settings: A Scoping Review},
  ISSN = {2168-8184},
  url = {http://dx.doi.org/10.7759/cureus.73994},
  DOI = {10.7759/cureus.73994},
  journal = {Cureus},
  publisher = {Springer Science and Business Media LLC},
  author = {Lee,  Craig and Britto,  Shawn and Diwan,  Khaled},
  year = {2024},
  month = nov 
}

@article{Tierney2024,
  title = {Ambient Artificial Intelligence Scribes to Alleviate the Burden of Clinical Documentation},
  volume = {5},
  ISSN = {2642-0007},
  url = {http://dx.doi.org/10.1056/CAT.23.0404},
  DOI = {10.1056/cat.23.0404},
  number = {3},
  journal = {NEJM Catalyst},
  publisher = {Massachusetts Medical Society},
  author = {Tierney,  Aaron A. and Gayre,  Gregg and Hoberman,  Brian and Mattern,  Britt and Ballesca,  Manuel and Kipnis,  Patricia and Liu,  Vincent and Lee,  Kristine},
  year = {2024},
  month = feb 
}

@article{Workman2023,
  title = {Identifying suicide documentation in clinical notes through zero‐shot learning},
  volume = {6},
  ISSN = {2398-8835},
  url = {http://dx.doi.org/10.1002/hsr2.1526},
  DOI = {10.1002/hsr2.1526},
  number = {9},
  journal = {Health Science Reports},
  publisher = {Wiley},
  author = {Workman,  Terri Elizabeth and Goulet,  Joseph L. and Brandt,  Cynthia A. and Warren,  Allison R. and Eleazer,  Jacob and Skanderson,  Melissa and Lindemann,  Luke and Blosnich,  John R. and O’Leary,  John and Zeng‐Treitler,  Qing},
  year = {2023},
  month = sep 
}

@article{Edgcomb2024,
  title = {Electronic Health Record Phenotyping of Pediatric Suicide-Related Emergency Department Visits},
  volume = {7},
  ISSN = {2574-3805},
  url = {http://dx.doi.org/10.1001/jamanetworkopen.2024.42091},
  DOI = {10.1001/jamanetworkopen.2024.42091},
  number = {10},
  journal = {JAMA Network Open},
  publisher = {American Medical Association (AMA)},
  author = {Edgcomb,  Juliet Beni and Olde Loohuis,  Loes and Tseng,  Chi-hong and Klomhaus,  Alexandra M. and Choi,  Kristen R. and Ponce,  Chrislie G. and Zima,  Bonnie T.},
  year = {2024},
  month = oct,
  pages = {e2442091}
}

@article{Liu2025,
author = {Liu, Di and Bai, Jingwen and Zhang, Zhuoyi and Zhang, Yilin and Zhang, Zhenhao and Zhao, Jian and An, Pengcheng},
title = {TherAIssist: Assisting Art Therapy Homework and Client-Practitioner Collaboration through Human-AI Interaction},
year = {2025},
issue_date = {September 2025},
publisher = {Association for Computing Machinery},
address = {New York, NY, USA},
volume = {9},
number = {3},
url = {https://doi.org/10.1145/3749497},
doi = {10.1145/3749497},
journal = {Proc. ACM Interact. Mob. Wearable Ubiquitous Technol.},
month = sep,
articleno = {113},
numpages = {38}
}

@article{Yuan2025,
    author = {Yuan, Aijia and Garcia Colato, Edlin and Pescosolido, Bernice and Song, Hyunju and Samtani, Sagar},
    title = {Improving Workplace Well-being in Modern Organizations: A Review of Large Language Model-based Mental Health Chatbots},
    year = {2025},
    issue_date = {March 2025},
    publisher = {Association for Computing Machinery},
    address = {New York, NY, USA},
    volume = {16},
    number = {1},
    issn = {2158-656X},
    url = {https://doi.org/10.1145/3701041},
    doi = {10.1145/3701041},
    journal = {ACM Trans. Manage. Inf. Syst.},
    month = feb,
    articleno = {3},
    numpages = {26}
}

@article{Kazantzis2018,
  title = {The Processes of Cognitive Behavioral Therapy: A Review of Meta-Analyses},
  volume = {42},
  ISSN = {1573-2819},
  url = {http://dx.doi.org/10.1007/s10608-018-9920-y},
  DOI = {10.1007/s10608-018-9920-y},
  number = {4},
  journal = {Cognitive Therapy and Research},
  publisher = {Springer Science and Business Media LLC},
  author = {Kazantzis,  Nikolaos and Luong,  Hoang Kim and Usatoff,  Alexsandra S. and Impala,  Tara and Yew,  Rui Ying and Hofmann,  Stefan G.},
  year = {2018},
  month = may,
  pages = {349–357}
}

@article{Westra2007,
  title = {Expectancy,  homework compliance,  and initial change in cognitive-behavioral therapy for anxiety.},
  volume = {75},
  ISSN = {0022-006X},
  url = {http://dx.doi.org/10.1037/0022-006X.75.3.363},
  DOI = {10.1037/0022-006x.75.3.363},
  number = {3},
  journal = {Journal of Consulting and Clinical Psychology},
  publisher = {American Psychological Association (APA)},
  author = {Westra,  Henny A. and Dozois,  David J. A. and Marcus,  Madalyn},
  year = {2007},
  month = jun,
  pages = {363–373}
}

@article{Heinz2025,
  title = {Randomized Trial of a Generative AI Chatbot for Mental Health Treatment},
  volume = {2},
  ISSN = {2836-9386},
  url = {http://dx.doi.org/10.1056/AIoa2400802},
  DOI = {10.1056/aioa2400802},
  number = {4},
  journal = {NEJM AI},
  publisher = {Massachusetts Medical Society},
  author = {Heinz,  Michael V. and Mackin,  Daniel M. and Trudeau,  Brianna M. and Bhattacharya,  Sukanya and Wang,  Yinzhou and Banta,  Haley A. and Jewett,  Abi D. and Salzhauer,  Abigail J. and Griffin,  Tess Z. and Jacobson,  Nicholas C.},
  year = {2025},
  month = mar 
}

@article{Held2024,
  title = {A Novel Cognitive Behavioral Therapy–Based Generative AI Tool (Socrates 2.0) to Facilitate Socratic Dialogue: Protocol for a Mixed Methods Feasibility Study},
  volume = {13},
  ISSN = {1929-0748},
  url = {http://dx.doi.org/10.2196/58195},
  DOI = {10.2196/58195},
  journal = {JMIR Research Protocols},
  publisher = {JMIR Publications Inc.},
  author = {Held,  Philip and Pridgen,  Sarah A and Chen,  Yaozhong and Akhtar,  Zuhaib and Amin,  Darpan and Pohorence,  Sean},
  year = {2024},
  month = oct,
  pages = {e58195}
}

@article{FrankeFyen2025,
  title = {Artificial intelligence vs. human expert: Licensed mental health clinicians’ blinded evaluation of AI-generated and expert psychological advice on quality,  empathy,  and perceived authorship},
  volume = {41},
  ISSN = {2214-7829},
  url = {http://dx.doi.org/10.1016/j.invent.2025.100841},
  DOI = {10.1016/j.invent.2025.100841},
  journal = {Internet Interventions},
  publisher = {Elsevier BV},
  author = {Franke F\"{o}yen,  Ludwig and Zapel,  Emma and Lekander,  Mats and Hedman-Lagerl\"{o}f,  Erik and Linds\"{a}ter,  Elin},
  year = {2025},
  month = sep,
  pages = {100841}
}

@inproceedings{grieg2019visual,
  author    = {Grieg, N. A. and Lillehaug, S.-I. and Lamo, Y.},
  title     = {A Visual Analytics Dashboard to Support iCBT Therapists},
  booktitle = {Proceedings of the 17th Scandinavian Conference on Health Informatics},
  volume    = {161},
  pages     = {134--140},
  year      = {2019},
  publisher = {Linköping University Electronic Press},
  url       = {http://www.ep.liu.se/ecp/161/023/ecp19161023.pdf}
}

@article{Gondek2016,
  title = {Feedback from Outcome Measures and Treatment Effectiveness,  Treatment Efficiency,  and Collaborative Practice: A Systematic Review},
  volume = {43},
  ISSN = {1573-3289},
  url = {http://dx.doi.org/10.1007/s10488-015-0710-5},
  DOI = {10.1007/s10488-015-0710-5},
  number = {3},
  journal = {Administration and Policy in Mental Health and Mental Health Services Research},
  publisher = {Springer Science and Business Media LLC},
  author = {Gondek,  Dawid and Edbrooke-Childs,  Julian and Fink,  Elian and Deighton,  Jessica and Wolpert,  Miranda},
  year = {2016},
  month = jan,
  pages = {325–343}
}

@article{Reiter2022,
  title = {Collaborative Documentation: Therapist Experiences in Jointly Writing Progress Notes},
  volume = {41},
  ISSN = {1195-4396},
  url = {http://dx.doi.org/10.1521/jsyt.2022.41.2.1},
  DOI = {10.1521/jsyt.2022.41.2.1},
  number = {2},
  journal = {Journal of Systemic Therapies},
  publisher = {Guilford Publications},
  author = {Reiter,  Michael D. and Bibliowicz,  Vanessa and Sabo,  Kayleigh and Yu,  Xinyan Cindy and Delgado,  Yesenia and Barrionuevo,  Desiree and Rich,  Bailey},
  year = {2022},
  month = jun,
  pages = {1–16}
}

@article{Lattie2025,
  title = {Examining the Client Experience of Digital Tools in Blended Care Therapy: Qualitative Interview Study},
  volume = {9},
  ISSN = {2561-326X},
  url = {http://dx.doi.org/10.2196/68249},
  DOI = {10.2196/68249},
  journal = {JMIR Formative Research},
  publisher = {JMIR Publications Inc.},
  author = {Lattie,  Emily G and Beltzer,  Miranda and Varra,  Alethea and Chen,  Connie E and Lungu,  Anita},
  year = {2025},
  month = apr,
  pages = {e68249}
}

@article{Wentzel2016,
  title = {Mixing Online and Face-to-Face Therapy: How to Benefit From Blended Care in Mental Health Care},
  volume = {3},
  ISSN = {2368-7959},
  url = {http://dx.doi.org/10.2196/mental.4534},
  DOI = {10.2196/mental.4534},
  number = {1},
  journal = {JMIR Mental Health},
  publisher = {JMIR Publications Inc.},
  author = {Wentzel,  Jobke and van der Vaart,  Rosalie and Bohlmeijer,  Ernst T and van Gemert-Pijnen,  Julia E W C},
  year = {2016},
  month = feb,
  pages = {e9}
}

@article{Kuhn2014,
  title = {Preliminary Evaluation of PTSD Coach,  a Smartphone App for Post-Traumatic Stress Symptoms},
  volume = {179},
  ISSN = {1930-613X},
  url = {http://dx.doi.org/10.7205/MILMED-D-13-00271},
  DOI = {10.7205/milmed-d-13-00271},
  number = {1},
  journal = {Military Medicine},
  publisher = {Oxford University Press (OUP)},
  author = {Kuhn,  Eric and Greene,  Carolyn and Hoffman,  Julia and Nguyen,  Tam and Wald,  Laura and Schmidt,  Janet and Ramsey,  Kelly M. and Ruzek,  Josef},
  year = {2014},
  month = jan,
  pages = {12–18}
}

@article{Reger2013,
  title = {The “PE coach” smartphone application: An innovative approach to improving implementation,  fidelity,  and homework adherence during prolonged exposure.},
  volume = {10},
  ISSN = {1541-1559},
  url = {http://dx.doi.org/10.1037/a0032774},
  DOI = {10.1037/a0032774},
  number = {3},
  journal = {Psychological Services},
  publisher = {American Psychological Association (APA)},
  author = {Reger,  Greg M. and Hoffman,  Julia and Riggs,  David and Rothbaum,  Barbara O. and Ruzek,  Josef and Holloway,  Kevin M. and Kuhn,  Eric},
  year = {2013},
  month = aug,
  pages = {342–349}
}

@article{Owen2015,
  title = {mHealth in the Wild: Using Novel Data to Examine the Reach,  Use,  and Impact of PTSD Coach},
  volume = {2},
  ISSN = {2368-7959},
  url = {http://dx.doi.org/10.2196/mental.3935},
  DOI = {10.2196/mental.3935},
  number = {1},
  journal = {JMIR Mental Health},
  publisher = {JMIR Publications Inc.},
  author = {Owen,  Jason E and Jaworski,  Beth K and Kuhn,  Eric and Makin-Byrd,  Kerry N and Ramsey,  Kelly M and Hoffman,  Julia E},
  year = {2015},
  month = mar,
  pages = {e7}
}

@article{Kuhn2015,
  title = {Clinician characteristics and perceptions related to use of the PE (prolonged exposure) coach mobile app.},
  volume = {46},
  ISSN = {0735-7028},
  url = {http://dx.doi.org/10.1037/pro0000051},
  DOI = {10.1037/pro0000051},
  number = {6},
  journal = {Professional Psychology: Research and Practice},
  publisher = {American Psychological Association (APA)},
  author = {Kuhn,  Eric and Crowley,  Jill J. and Hoffman,  Julia E. and Eftekhari,  Afsoon and Ramsey,  Kelly M. and Owen,  Jason E. and Reger,  Greg M. and Ruzek,  Josef I.},
  year = {2015},
  month = dec,
  pages = {437–443}
}

@article{Kuo2022,
    author = {Kuo, Patty Beyrong and Soma, Christina and Axford, Katherine and Hirsch, Tad and Imel, Zac E. and Van Epps, Jake J.},
    title = {Do as I say, not as I do: Therapist Evaluation of a Practice and Supervision Aid},
    year = {2022},
    issue_date = {November 2022},
    publisher = {Association for Computing Machinery},
    address = {New York, NY, USA},
    volume = {6},
    number = {CSCW2},
    url = {https://doi.org/10.1145/3555185},
    doi = {10.1145/3555185},
    journal = {Proc. ACM Hum.-Comput. Interact.},
    month = nov,
    articleno = {294},
    numpages = {23}
}

@article{Oewel2024a,
    author = {Oewel, Bruna and Azizan, Nadia and Arean, Patricia A. and Agapie, Elena},
    title = {Technology's Role in Fostering Therapist-Client Collaboration and Engagement with Goals},
    year = {2024},
    issue_date = {November 2024},
    publisher = {Association for Computing Machinery},
    address = {New York, NY, USA},
    volume = {8},
    number = {CSCW2},
    url = {https://doi.org/10.1145/3687055},
    doi = {10.1145/3687055},
    journal = {Proc. ACM Hum.-Comput. Interact.},
    month = nov,
    articleno = {516},
    numpages = {28}
}

@inproceedings{Oewel2024b,
    author = {Oewel, Bruna and Arean, Patricia Anne and Agapie, Elena},
    title = {Approaches for tailoring between-session mental health therapy activities},
    year = {2024},
    isbn = {9798400703300},
    publisher = {Association for Computing Machinery},
    address = {New York, NY, USA},
    url = {https://doi.org/10.1145/3613904.3642856},
    doi = {10.1145/3613904.3642856},
    booktitle = {Proceedings of the 2024 CHI Conference on Human Factors in Computing Systems},
    articleno = {696},
    numpages = {19},
    location = {Honolulu, HI, USA},
    series = {CHI '24}
}

@inproceedings{Stawarz2020,
    author = {Stawarz, Katarzyna and Preist, Chris and Tallon, Deborah and Thomas, Laura and Turner, Katrina and Wiles, Nicola and Kessler, David and Shafran, Roz and Coyle, David},
    title = {Integrating the Digital and the Traditional to Deliver Therapy for Depression: Lessons from a Pragmatic Study},
    year = {2020},
    isbn = {9781450367080},
    publisher = {Association for Computing Machinery},
    address = {New York, NY, USA},
    url = {https://doi.org/10.1145/3313831.3376510},
    doi = {10.1145/3313831.3376510},
    booktitle = {Proceedings of the 2020 CHI Conference on Human Factors in Computing Systems},
    pages = {1–14},
    numpages = {14},
    location = {Honolulu, HI, USA},
    series = {CHI '20}
}

@article{Bhat2021,
    author = {Bhat, Ashwin Sadananda and Boersma, Christiaan and Meijer, Max Jan and Dokter, Maaike and Bohlmeijer, Ernst and Li, Jamy},
    title = {Plant Robot for At-Home Behavioral Activation Therapy Reminders to Young Adults with Depression},
    year = {2021},
    issue_date = {September 2021},
    publisher = {Association for Computing Machinery},
    address = {New York, NY, USA},
    volume = {10},
    number = {3},
    url = {https://doi.org/10.1145/3442680},
    doi = {10.1145/3442680},
    month = jul,
    articleno = {20},
    numpages = {21}
}

@article{Evans2024,
    author = {Evans, Hayley I. and Ryu, Myeonghan and Hsieh, Theresa and Zhou, Jiawei and Xu, Kefan and Akers, Kenneth W. and Sherrill, Andrew M. and Arriaga, Rosa I.},
    title = {Using Sensor-Captured Patient-Generated Data to Support Clinical Decision-making in PTSD Therapy},
    year = {2024},
    issue_date = {April 2024},
    publisher = {Association for Computing Machinery},
    address = {New York, NY, USA},
    volume = {8},
    number = {CSCW1},
    url = {https://doi.org/10.1145/3637426},
    doi = {10.1145/3637426},
    journal = {Proc. ACM Hum.-Comput. Interact.},
    month = apr,
    articleno = {149},
    numpages = {28}
}

@article{Bunnell2021,
  title        = {Barriers associated with the implementation of homework in youth mental health treatment and potential mobile health solutions},
  author       = {Bunnell, Brian E. and Nemeth, Leona S. and Lenert, Lisa A. and Kazantzis, Nikolaos and Deblinger, Esther and Higgins, Kyle A. and Ruggiero, Kenneth J.},
  journal      = {Cognitive Therapy and Research},
  year         = {2021},
  volume       = {45},
  pages        = {272-286},
  doi          = {10.1007/s10608-020-10090-8},
  url          = {https://doi.org/10.1007/s10608-020-10090-8}
}

@article{Kazantzis2002,
  title = {Reflecting on homework in psychotherapy: What can we conclude from research and experience?},
  volume = {58},
  ISSN = {1097-4679},
  url = {http://dx.doi.org/10.1002/jclp.10034},
  DOI = {10.1002/jclp.10034},
  number = {5},
  journal = {Journal of Clinical Psychology},
  publisher = {Wiley},
  author = {Kazantzis,  Nikolaos and Lampropoulos,  Georgios K.},
  year = {2002},
  month = apr,
  pages = {577–585}
}

@article{maniss2018collaborative,
  title = {Collaborative Documentation for Behavioral Healthcare Providers: An Emerging Practice},
  author = {Maniss, Suzanne and Pruit, Amanda G.},
  journal = {Journal of Human Services: Training, Research, and Practice},
  volume = {3},
  number = {1},
  eid = {2},
  year = {2018},
  url = {https://scholarworks.sfasu.edu/jhstrp/vol3/iss1/2}
}

@article{Asgari2025,
  title = {A framework to assess clinical safety and hallucination rates of LLMs for medical text summarisation},
  volume = {8},
  ISSN = {2398-6352},
  url = {http://dx.doi.org/10.1038/s41746-025-01670-7},
  DOI = {10.1038/s41746-025-01670-7},
  number = {1},
  journal = {npj Digital Medicine},
  publisher = {Springer Science and Business Media LLC},
  author = {Asgari,  Elham and Montaña-Brown,  Nina and Dubois,  Magda and Khalil,  Saleh and Balloch,  Jasmine and Yeung,  Joshua Au and Pimenta,  Dominic},
  year = {2025},
  month = may 
}

@article{Topaz2025,
  title = {Beyond human ears: navigating the uncharted risks of AI scribes in clinical practice},
  volume = {8},
  ISSN = {2398-6352},
  url = {http://dx.doi.org/10.1038/s41746-025-01895-6},
  DOI = {10.1038/s41746-025-01895-6},
  number = {1},
  journal = {npj Digital Medicine},
  publisher = {Springer Science and Business Media LLC},
  author = {Topaz,  Maxim and Peltonen,  Laura Maria and Zhang,  Zhihong},
  year = {2025},
  month = sep 
}

@inproceedings{Zhang2024,
author = {Zhang, Yixuan and Wang, Yimeng and Yongsatianchot, Nutchanon and Gaggiano, Joseph D and Suhaimi, Nurul M and Okrah, Anne and Kim, Miso and Griffin, Jacqueline and Parker, Andrea G},
title = {Profiling the Dynamics of Trust \& Distrust in Social Media: A Survey Study},
year = {2024},
isbn = {9798400703300},
publisher = {Association for Computing Machinery},
address = {New York, NY, USA},
url = {https://doi.org/10.1145/3613904.3642927},
doi = {10.1145/3613904.3642927},
booktitle = {Proceedings of the 2024 CHI Conference on Human Factors in Computing Systems},
articleno = {947},
numpages = {24},
location = {Honolulu, HI, USA},
series = {CHI '24}
}

\appendix
\section{Appendices}
\label{sec:appendices}

\subsection{Survey Questions}
\label{sec:survey_questions}
\textbf{Current Practices of Assigning and Tracking Homework:}

1) \textit{How many years have you been providing therapy?} (Numeric numbers are allowed.)

2) \textit{How do you currently track your clients' completion of homework?} (Participants were allowed to select all applicable options from the following list: ``Client self-report,'' ``Technology-aided tools (e.g., web apps, mobile apps),'' ``Worksheet completion (paper-based),'' ``Other.'')

3) \textit{Which types of homework do you typically assign?} (Participants were allowed to select all applicable options from the following list: ``Activity scheduling,'' ``Behavioral experiments,'' ``Cognitive restructuring exercises,'' ``Exposure tasks,'' ``Journaling,'' ``Mindfulness practice,'' ``Thought records,'' ``Other.'')

\textbf{Tracked Homework Details and Associated Challenges:}

1) \textit{Please describe the specific details you ask clients to record for each type of homework and the purpose of asking for each detail. We’re interested in understanding not only what you instruct them to track, but also why these details are important in providing you with insights as their therapist. For clarity, please list each detail separately and explain the value it brings to the therapeutic process.} (Participants responded to an open-ended question without any restrictions.)

2) \textit{What are your current pain points in tracking clients' completion and progress related to homework?} (Participants responded to an open-ended question without any restrictions.)

\textbf{Desired Features for Technology-Assisted Support:}

1) \textit{If you had a ``magic wand'', which of the following features would you like to have to help you track clients' completion of homework and progress using technology-aided solutions? } (Participants were allowed to select all applicable options from the following list: ``Ability to provide suggestions on next homework assignments based on client progress,'' ``Client’s completion rates of specific exercises,'' ``Summaries of client progress in completing specific exercises,'' ``Tracking changes in client condition (e.g., self-ratings of emotions like SUDS, self-efficacy),'' ``Access to clients' completed exercises for review,'' ``Time spent on specific exercises,'' ``Quality of client response,'' ``Client's self-reported insights,'' ``Visual presentation of clients' progress data'', ``Other.'')

2) \textit{Which aspects of the homework do you believe would benefit from technological assistance?} (Participants responded to an open-ended question without any restrictions.)

\subsection{Likert Scale for Pilot Study}
\label{sec:likert}
\textbf{Perceived Ease of Use:}

1) \textit{The dashboard is easy to navigate.}

\textit{Rationale:} Captures the overall intuitiveness of the system layout, a key factor in minimizing effort and cognitive load during clinical use.

2) \textit{The graphs and metrics on the dashboard are visually clear.}

\textit{Rationale:} Reflects how easily visualized data can be interpreted. Readability of visual output is crucial for workflow efficiency.

\textbf{Perceived Usefulness:}

1) \textit{The AI-generated summaries help reduce my workload.} 

\textit{Rationale:} Gauges whether the AI component meaningfully reduces therapists’ administrative burden.

2) \textit{I can efficiently find the information I need on the dashboard.} 

\textit{Rationale:} Assesses whether \tool helps therapists locate and use relevant information quickly to support clinical decision-making.

3) \textit{The dashboard effectively supports my ability to assess and track patient homework.} 

\textit{Rationale:} Directly evaluates the system’s relevance to a core therapeutic task: understanding client engagement between sessions.

\textbf{Impact:}

1) \textit{I am likely to integrate this tool into my practice if it becomes available long-term.}

\textit{Rationale:} Measures behavioral intention to adopt, indicating whether the system could influence practice over time.

2) \textit{I would recommend this dashboard to other therapists. }

\textit{Rationale:} Reflects overall satisfaction and perceived value beyond personal use.

\textbf{User Control:}

1) \textit{The system allows me to maintain sufficient control over its automated features.}

\textit{Rationale:} Evaluating the balance between manual control and AI automation helps capture how well the system respects clinicians’ need for control while leveraging AI assistance.

\textbf{Trust:}

1) \textit{I trust the AI summary and assistant.}

\textit{Rationale: Trust captures clinicians’ confidence in the reliability and appropriateness of AI-generated outputs, which is essential for meaningful engagement.}

\subsection{Simulated Client Information}

\hl{Below, we provide example details of the simulated client, ``Elias'', including demographics, clinical background, and sample data logs used in the study.}

\subsubsection{Client Clinical Background}

Elias is a first-year PhD student experiencing significant stress related to his transition into graduate school. He reports fear of negative evaluation from his advisor and catastrophic thinking regarding his career prospects. He tracks his biohealth data, emotions, and daily activities. He completes therapeutic homework 2-3 times a week.

\subsubsection{Homework Data}

\autoref{tab:journals} displays a subset of the simulated Thought Records Homework generated for the study.

\begin{table*}[h]
    \centering
    \footnotesize
    \caption{Simulated Thought Record Homework Entries reflecting stressors.}
    \begin{tabularx}{\textwidth}{p{0.2\textwidth} p{0.3\textwidth} X}
        \toprule
        \textbf{Trigger Event} & \textbf{Automatic Negative Thought} & \textbf{Rational Response / New Thought} \\
        \midrule
        My paper got rejected. & My papers will never be accepted. & I am in the first year of my PhD program, and I still have a lot of time; take this rejection as a learning experience, and I will produce high-quality articles. \\
        \midrule
        Students send endless emails after office hours. & I want to quit my job. & I can understand why they do that; I was a student once, and I probably did something similar to bother my TA. \\
        \midrule
        I called my girlfriend, but she did not answer. & She is ignoring my call intentionally. & She is probably focusing on her work and just hasn't checked her phone yet. \\
        \midrule
        My best friend did not reply to my SMS. & We are not best friends anymore. & Since we are best friends, he probably thinks it is okay to delay responding to my message, which implies our relationship is secure. \\
        \midrule
        Reading a very complex and technical manuscript. & I am not able to re-implement their idea. & Probably not, but I can try step-by-step. I am not required to re-implement the whole thing, so if I implement the majority of it, it is a success. \\
        \midrule
        Talking to ChatGPT. & AI will replace me very soon. & This is probably true, but I can't do anything about it, so why worry? \\
        \midrule
        Watching political news. & This world sucks. & I think good people are in the majority and there are so many good things happening in this world; only politics sucks. \\
        \bottomrule
    \end{tabularx}
    \label{tab:journals}
\end{table*}

\subsubsection{Emotion and Activity Logs}

We simulated longitudinal data covering daily mood and activities. \textbf{Emotion} was recorded in 4-hour intervals (e.g., 6am--10am, 10am--2pm, etc.) throughout the client's waking hours. The dataset utilizes descriptors such as \textit{Energetic, Overwhelmed, Sleepy, Enthusiastic, Bored,} and \textit{Relaxed} to reflect fluctuations in the client's emotional state. \textbf{Activity} was generated for three daily time blocks (Morning, Afternoon, Night). These entries capture a diverse range of behaviors, including physical exercise (e.g., swimming, running), social interactions (e.g., calling parents, dining with friends), and solitary hobbies (e.g., baking, gaming).

\subsection{Prompt Examples}
\label{sec:prompt_examples}

\subsubsection{Example 1: Chat Assistant Prompt}
\label{sec:ex_chatbot}

\begin{Verbatim}[
  fontsize=\footnotesize,
  breaklines=true, 
  breaksymbol={},  
  commandchars=\\\{\}  
]
You are a clinical-support assistant that answers therapist questions about a client’s psychotherapy homework data. Your role is to generate concise, evidence-linked responses strictly based on the provided context. Do not infer beyond the inputs or provide clinical advice. Your answers must support therapists in interpreting patterns, comparisons, or risks, while preserving transparency and auditability.

INPUT
You will receives two types of information:
1. Parameters:
   - question: the therapist’s natural-language query
   - question_category (e.g., journaling, homework, biometric, risk, suggestion, general)
   - question_scope (i.e., recent, comparative)
   - focus_areas: therapist-defined emphasis areas
   - aiChatAbilities: preferences indicating whether raw data extraction and/or detailed explanations should be included
2. Client data:
   - homework_data: structured per-date submissions
   - biometric_aggregates: sleep, heart rate, activity metrics
   - reading_materials: completion status

PROCESS
1. Scope check
- If question_scope = "recent": use only the most recent "to_now" data.
- If "comparative": highlight changes across time spans (e.g., week-over-week).
2. Raw data option
- If aiChatAbilities includes a raw-data preference, add a block titled “Relevant Raw Data” listing only the directly relevant data points. 
- Do not interpret or summarize this block; simply present the data snippets.
- If no such preference, omit this block.
3. Therapist focus
- Emphasize evidence related to focus_areas and homework_types wherever possible.
- If absent, state “No data related to focus areas”.
4. Category-specific instructions
- journaling: extract recurring themes, emotions, or thought patterns from client entries.
- homework: describe completion trends, consistency, gaps, or streaks.
- biometric: summarize biometric patterns (sleep, heart rate, activity), anchoring metrics.
- risk: list potential warning signs of emotional distress, using cautious phrasing (e.g., “Possible signals include...”) and concrete anchors (what/when/source).
- suggestion: Begin with the line: [Disclaimer] AI-generated suggestions may be incomplete or contain errors. Review raw client data before acting. Then provide less than 6 actionable bullet points, each linked to evidence from the client data.
- general: extract only the minimal details that directly answer the question.


HARD CONSTRAINTS
- If no relevant evidence exists, output exactly “Insufficient data”.
- Responses must be in plain text, using concise bullet points.
- Do not fabricate numbers, events, or any content not present in the inputs.
- Report conflicting evidence neutrally (e.g., “Conflicting entries observed:...”); do not attempt to resolve.
- Maintain neutral, concise, evidence-linked phrasing without hedging or repetition.
- Avoid meta-commentary and do not reference these instructions.


\end{Verbatim}

\subsubsection{Example 2: GenAI Summary Prompt}
\label{sec:ex_summary}

\begin{Verbatim}[
  fontsize=\footnotesize,
  breaklines=true, 
  breaksymbol={},  
  commandchars=\\\{\}  
]
\textbf{System Prompt}
You are a clinical-support summarizer for psychotherapy homework review only. You should ensure summaries remain faithful to the source data, and provide therapists with concise trend insights, avoiding hidden inferences or reliance on external knowledge. Structure output with clear, well-defined sections to support consistent downstream parsing and transparent human review.

INPUT
You will receives two types of information:
1. Parameters that specify how the summary should be generated:
summary_level (e.g., Basic Overview, Detailed Analysis, or No AI Summary)
summary_priorities (selected focus categories)
homework_summary (daily, weekly, or none)
focus_areas (list of emphasis areas)
homework_types (types of assigned homework)
2. Data extracted from the JSON user record:
reading_materials: { finished: [...], not_finished: [...] }
biometric_aggregates: averages/ranges text block, or “No data”
homework_data: structured per-date text

PROCESS
1. Scope check
- Use ONLY the provided inputs. If a detail is absent, write "No data".
- If summary_level = "No AI Summary": output exactly "No AI summary is needed." and stop.
2. Priority routing
- Address summary_priorities FIRST; place remaining domains under "Other observations".
- Applies to both Basic Overview and Detailed Analysis.
3. Therapist focus
- Highlight evidence relevant to focus_areas and homework_types throughout the summary.
- Applies to both Basic Overview and Detailed Analysis.
4. Frequency control
- homework_summary = "daily": aggregate by day; do not copy raw submissions.
- homework_summary = "weekly": emphasize week-level patterns and changes.
- homework_summary = "none": only state whether homework was submitted.
- Applies to both Basic Overview and Detailed Analysis.
5. Granularity control
- Basic Overview: Write one plain-text paragraph covering reading materials, homework completion, mental health patterns, journaling/thought records, biometric data (if any), and risk signals. Mention each only at a high level; do not use headings.
- Detailed Analysis: Write a structured report using the section-specific instructions below.

INSTRUCTIONS (for Detailed Analysis)
- Reading Materials: State which were completed and which remain unfinished.
- Homework Completion Trends: Summarize overall patterns, highlighting consistency, gaps, and variation over time. Integrate evidence from focus_areas and homework_types, and explicitly connect these trends to therapist-defined goals (e.g., whether progress is evident in emotion regulation).
- Mental Health Patterns: Identify recurring emotional themes, triggers, or behavioral shifts visible in submissions, prioritizing evidence tied to therapist-defined focus areas.
- Journaling & Thought Records: Highlight the frequency and depth of entries, extract recurring cognitive patterns, and assess completeness of thought record fields (situation–automatic thought–evidence–reframe).
- Biometric Trends from Apple Health: Summarize sleep, heart rate, and activity patterns, and, where relevant, relate changes to adherence or stress.
- Risk Alerts for Emotional Distress: Identify potential warning signs if signals are present; otherwise state "No data". Use calibrated language (e.g., "Possible"; "Signals include...") and attach concrete anchors (what/when/source).
- Other Observations: Include only if relevant evidence is present for non-priority areas.
- Overall Summary: Conclude with 1–2 sentences synthesizing the client’s progress in a neutral, concise, evidence-based tone.
- Output must be plain text, with section headers presented in the following exact order: Reading Materials, Homework Completion Trends, Mental Health Patterns, Journaling & Thought Records, Biometric Trends from Apple Health, Risk Alerts for Emotional Distress, Overall Summary.

HARD CONSTRAINTS
- Always include the section header followed by “No data” if information is missing.
- Do not provide recommendations, diagnoses, treatments, probabilities, or external knowledge.
- Do not output raw data, tables, or headings beyond those specified.
- Do not fabricate numbers, events, or content not present in the inputs.
- Report conflicting data neutrally (e.g., “Conflicting entries observed:...”); do not attempt to resolve them.
- Maintain neutral, concise, evidence-linked phrasing without hedging or repetition.
- Avoid meta-commentary and do not reference these instructions.

\end{Verbatim}

\end{document}